\documentclass{article}
\pdfoutput=1 
\usepackage{arxiv}

\usepackage[utf8]{inputenc} 
\usepackage[T1]{fontenc}    
\usepackage{lmodern}        
\usepackage{hyperref}       
\usepackage{url}            
\usepackage{booktabs}       
\usepackage{amsfonts}       
\usepackage{nicefrac}       
\usepackage{microtype}      
\usepackage{graphicx}

\title{Statistical optimization of expensive multi-response black-box functions}

\author{
    Andreas Mändle
    \thanks{Corresponding author}
   \\
    Department for Modelling and Simulation \\
    Faserinstitut Bremen e.V. \\
  Am Biologischen Garten 2, 28359 Bremen \\
  \texttt{\href{mailto:maendle@uni-bremen.de}{\nolinkurl{maendle@uni-bremen.de}}} \\
   \And
    Werner Brannath
   \\
    Institute for Statistics \\
    University of Bremen \\
  Linzer Straße 4, 28359 Bremen \\
  \texttt{\href{mailto:wwosniok@math.uni-bremen.de}{\nolinkurl{wwosniok@math.uni-bremen.de}}} \\
   \And
    Werner Wosniok
   \\
    Institute for Statistics \\
    University of Bremen \\
  Linzer Straße 4, 28359 Bremen \\
  \texttt{\href{mailto:wwosniok@math.uni-bremen.de}{\nolinkurl{wwosniok@math.uni-bremen.de}}} \\
  }

\usepackage{longtable,booktabs,array}
\usepackage{calc} 
\usepackage{etoolbox}
\makeatletter
\patchcmd\longtable{\par}{\if@noskipsec\mbox{}\fi\par}{}{}
\makeatother
\IfFileExists{footnotehyper.sty}{\usepackage{footnotehyper}}{\usepackage{footnote}}
\makesavenoteenv{longtable}

\newlength{\cslhangindent}
\newlength{\csllabelwidth}
\newlength{\cslentryspacingunit} 
  {}%
  {\par}
\newenvironment{CSLReferences}[2] 
 {
  \setlength{\parindent}{0pt}
  \ifodd #1
  \let\oldpar\par
  \def\par{\hangindent=\cslhangindent\oldpar}
  \fi
  \setlength{\parskip}{#2\cslentryspacingunit}
 }%
 {}
\usepackage{calc}

\usepackage{tikz}
\usepackage{multicol}
\usepackage{bbm}
\usepackage{longtable}
\usepackage{booktabs}
\usepackage{grffile}
\usepackage{hyperref}
\usepackage{calc}
\usepackage{placeins}
\usepackage{float}
\usepackage{pdflscape}
\usepackage{placeins}
\usepackage{amsmath}
\usetikzlibrary{patterns}
\begin{document}
\maketitle

\begin{abstract}
Assume that a set of \(P\) process parameters \(p_i\), \(i=1,\dots,P\), determines the outcome of a set of \(D\) descriptor variables \(d_j\), \(j=1,\dots,D\), via an unknown functional relationship \(\phi: \mathbf{p} \mapsto \mathbf{d}, \, \mathbb{R}^{P} \to \mathbb{R}^{D}\), where \(\mathbf{p}=(p_1,\dots,p_{P})\), \(\mathbf{d}=(d_1,\dots,d_{D})\). It is desired to find appropriate values \(\mathbf{\hat p} = ({\hat p}_1,\dots, {\hat p}_P)\) for the process parameters such that the corresponding values of the descriptor variables \(\phi (\mathbf {\hat p})\) are close to a given target \(\mathbf d^*=(d^*_1,\dots,d^*_D)\), assuming that at least one exact solution exists. A sequential approach using dimension reduction techniques has been developed to achieve this. In a simulation study, results of the suggested approach and the algorithms NSGA-II, SMS-EMOA and MOEA/D are compared.
\end{abstract}


\newcommand{\Li}[1][]{\Lambda_i^{#1}}
\newcommand{\thi}[1][] {\theta_i^{#1}}
\newcommand{\di}[2][i] {d_{#1}^{(#2)}}
\newcommand{\Gi}{G_i}
\newcommand{\Hi}{H_i}
\newcommand{\Ai}[1][] {A_i^{(#1)}}
\newcommand{\Aj}[1][] {A^{(#1)}}
\newcommand{\Bj}[1][] {B^{(#1)}}
\newcommand{\Bi}[1][] {B_i^{(#1)}}
\newcommand{\Bij}[2][i] {B_#1^{(#2)}}
\newcommand{\Aij}[2][i] {A_#1^{(#2)}}
\newcommand{\FWEi}{\operatorname{FWE}_I(\mu)}
\newcommand{\FWEii}{\operatorname{FWE}_{II}(\mu)}

\hypertarget{introduction}{%
\section{Introduction}\label{introduction}}

\begin{figure}[h]
\centering

 
\tikzset{
pattern size/.store in=\mcSize, 
pattern size = 5pt,
pattern thickness/.store in=\mcThickness, 
pattern thickness = 0.3pt,
pattern radius/.store in=\mcRadius, 
pattern radius = 1pt}
\makeatletter
\pgfutil@ifundefined{pgf@pattern@name@_a0bhw36ro}{
\pgfdeclarepatternformonly[\mcThickness,\mcSize]{_a0bhw36ro}
{\pgfqpoint{0pt}{0pt}}
{\pgfpoint{\mcSize}{\mcSize}}
{\pgfpoint{\mcSize}{\mcSize}}
{
\pgfsetcolor{\tikz@pattern@color}
\pgfsetlinewidth{\mcThickness}
\pgfpathmoveto{\pgfqpoint{0pt}{\mcSize}}
\pgfpathlineto{\pgfpoint{\mcSize+\mcThickness}{-\mcThickness}}
\pgfpathmoveto{\pgfqpoint{0pt}{0pt}}
\pgfpathlineto{\pgfpoint{\mcSize+\mcThickness}{\mcSize+\mcThickness}}
\pgfusepath{stroke}
}}
\makeatother

 
\tikzset{
pattern size/.store in=\mcSize, 
pattern size = 5pt,
pattern thickness/.store in=\mcThickness, 
pattern thickness = 0.3pt,
pattern radius/.store in=\mcRadius, 
pattern radius = 1pt}
\makeatletter
\pgfutil@ifundefined{pgf@pattern@name@_w94tfcoi9}{
\pgfdeclarepatternformonly[\mcThickness,\mcSize]{_w94tfcoi9}
{\pgfqpoint{0pt}{0pt}}
{\pgfpoint{\mcSize}{\mcSize}}
{\pgfpoint{\mcSize}{\mcSize}}
{
\pgfsetcolor{\tikz@pattern@color}
\pgfsetlinewidth{\mcThickness}
\pgfpathmoveto{\pgfqpoint{0pt}{\mcSize}}
\pgfpathlineto{\pgfpoint{\mcSize+\mcThickness}{-\mcThickness}}
\pgfpathmoveto{\pgfqpoint{0pt}{0pt}}
\pgfpathlineto{\pgfpoint{\mcSize+\mcThickness}{\mcSize+\mcThickness}}
\pgfusepath{stroke}
}}
\makeatother

 
\tikzset{
pattern size/.store in=\mcSize, 
pattern size = 5pt,
pattern thickness/.store in=\mcThickness, 
pattern thickness = 0.3pt,
pattern radius/.store in=\mcRadius, 
pattern radius = 1pt}
\makeatletter
\pgfutil@ifundefined{pgf@pattern@name@_r93r6ozlo}{
\pgfdeclarepatternformonly[\mcThickness,\mcSize]{_r93r6ozlo}
{\pgfqpoint{0pt}{0pt}}
{\pgfpoint{\mcSize}{\mcSize}}
{\pgfpoint{\mcSize}{\mcSize}}
{
\pgfsetcolor{\tikz@pattern@color}
\pgfsetlinewidth{\mcThickness}
\pgfpathmoveto{\pgfqpoint{0pt}{\mcSize}}
\pgfpathlineto{\pgfpoint{\mcSize+\mcThickness}{-\mcThickness}}
\pgfpathmoveto{\pgfqpoint{0pt}{0pt}}
\pgfpathlineto{\pgfpoint{\mcSize+\mcThickness}{\mcSize+\mcThickness}}
\pgfusepath{stroke}
}}
\makeatother
\tikzset{every picture/.style={line width=0.75pt}} 
\begin{tikzpicture}[x=0.75pt,y=0.75pt,yscale=-1,xscale=1]
\draw   (190.5,29.5) .. controls (190.5,23.98) and (194.98,19.5) .. (200.5,19.5) .. controls (206.02,19.5) and (210.5,23.98) .. (210.5,29.5) .. controls (210.5,35.02) and (206.02,39.5) .. (200.5,39.5) .. controls (194.98,39.5) and (190.5,35.02) .. (190.5,29.5) -- cycle ;
\draw   (175.5,51.5) .. controls (175.5,45.98) and (179.98,41.5) .. (185.5,41.5) .. controls (191.02,41.5) and (195.5,45.98) .. (195.5,51.5) .. controls (195.5,57.02) and (191.02,61.5) .. (185.5,61.5) .. controls (179.98,61.5) and (175.5,57.02) .. (175.5,51.5) -- cycle ;
\draw   (203.5,51.5) .. controls (203.5,45.98) and (207.98,41.5) .. (213.5,41.5) .. controls (219.02,41.5) and (223.5,45.98) .. (223.5,51.5) .. controls (223.5,57.02) and (219.02,61.5) .. (213.5,61.5) .. controls (207.98,61.5) and (203.5,57.02) .. (203.5,51.5) -- cycle ;
\draw  [pattern=_a0bhw36ro,pattern size=3.4499999999999997pt,pattern thickness=0.75pt,pattern radius=0pt, pattern color={rgb, 255:red, 0; green, 0; blue, 0}] (465,29.5) .. controls (465,23.98) and (469.48,19.5) .. (475,19.5) .. controls (480.52,19.5) and (485,23.98) .. (485,29.5) .. controls (485,35.02) and (480.52,39.5) .. (475,39.5) .. controls (469.48,39.5) and (465,35.02) .. (465,29.5) -- cycle ;
\draw  [pattern=_w94tfcoi9,pattern size=3.4499999999999997pt,pattern thickness=0.75pt,pattern radius=0pt, pattern color={rgb, 255:red, 0; green, 0; blue, 0}] (450,51.5) .. controls (450,45.98) and (454.48,41.5) .. (460,41.5) .. controls (465.52,41.5) and (470,45.98) .. (470,51.5) .. controls (470,57.02) and (465.52,61.5) .. (460,61.5) .. controls (454.48,61.5) and (450,57.02) .. (450,51.5) -- cycle ;
\draw  [pattern=_r93r6ozlo,pattern size=3.4499999999999997pt,pattern thickness=0.75pt,pattern radius=0pt, pattern color={rgb, 255:red, 0; green, 0; blue, 0}] (478,51.5) .. controls (478,45.98) and (482.48,41.5) .. (488,41.5) .. controls (493.52,41.5) and (498,45.98) .. (498,51.5) .. controls (498,57.02) and (493.52,61.5) .. (488,61.5) .. controls (482.48,61.5) and (478,57.02) .. (478,51.5) -- cycle ;
\draw  [dash pattern={on 4.5pt off 4.5pt}]  (458,56) -- (458.95,74.34) ;
\draw [shift={(459.1,77.33)}, rotate = 267.04] [fill={rgb, 255:red, 0; green, 0; blue, 0 }  ][line width=0.08]  [draw opacity=0] (6.25,-3) -- (0,0) -- (6.25,3) -- cycle    ;
\draw  [dash pattern={on 4.5pt off 4.5pt}]  (474.55,34) -- (475.58,74.67) ;
\draw [shift={(475.66,77.67)}, rotate = 268.55] [fill={rgb, 255:red, 0; green, 0; blue, 0 }  ][line width=0.08]  [draw opacity=0] (6.25,-3) -- (0,0) -- (6.25,3) -- cycle    ;
\draw  [dash pattern={on 4.5pt off 4.5pt}]  (488.9,56) -- (489.85,74.34) ;
\draw [shift={(490,77.33)}, rotate = 267.04] [fill={rgb, 255:red, 0; green, 0; blue, 0 }  ][line width=0.08]  [draw opacity=0] (6.25,-3) -- (0,0) -- (6.25,3) -- cycle    ;
\draw   (255,26.38) -- (394.02,26.38) -- (394.02,18) -- (405,35.67) -- (394.02,53.33) -- (394.02,44.95) -- (255,44.95) -- cycle ;
\draw (131,-0.5) node [anchor=north west][inner sep=0.75pt]  [font=\small] [align=left] {unprocessed samples};
\draw (414,-0.5) node [anchor=north west][inner sep=0.75pt]  [font=\small] [align=left] {processed material};
\draw (440.5,75) node [anchor=north west][inner sep=0.75pt]  [font=\small] [align=left] {$d_{1} ,\dotsc ,d_{D}$};
\draw (404,91) node [anchor=north west][inner sep=0.75pt]  [font=\small] [align=left] {\textbf{descriptor variables}};
\draw (293.5,29) node [anchor=north west][inner sep=0.75pt]  [font=\small] [align=left] {$p_{1} ,\dotsc ,p_{P}$};
\draw (257.5,52) node [anchor=north west][inner sep=0.75pt]  [font=\small] [align=left] {\textbf{predictor variables}};
\end{tikzpicture}
\caption{Predictors and descriptors for the treatment of homogeneous material samples.} \label{fig:idea}
\end{figure}
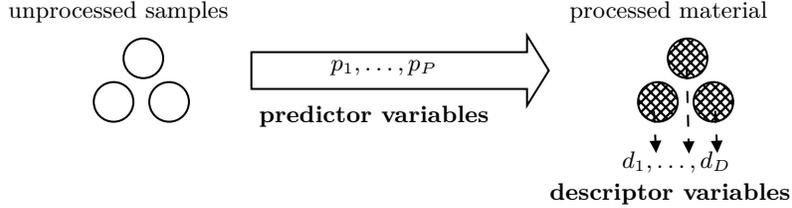

In the scope of the project \emph{CRC 1232 Farbige Zustände: High Throughput for Evolutionary Structural Materials}, cf.~
Ellendt and Mädler (2018), the relationship between multidimensional input and output parameters in the field of material science is investigated. Special interest lies in optimizing the input parameters, such that the output parameters are close to a target value. In a subdivision of the project the following setting is investigated, cf.~Figure \ref{fig:idea}:
a set of \(P\) micro process parameters (predictor variables) \(p_i\), \(i=1,\dots,P\) determines the outcome of \(D\) descriptor variables \(d_j\), \(j=1,\dots,D\) such that an unknown functional relationship \(\phi: \mathbf{p} \mapsto \mathbf{d},\,\mathbb R^P \to \mathbb R^D\) holds, where \(\mathbf{p}=(p_1,\dots, p_P)\) and \(\mathbf{d}=(d_1,\dots, d_D)\).

The aim is to find appropriate micro process parameters \({\hat p}_i\), \(i=1,\dots,P\) such that the corresponding micro descriptor variables \(\mathbf{d}=(d_1,\dots, d_D)\) are close to given target values \(\mathbf{d}^*=(d_1^*,\dots, d_D^*)\).
As \(\phi\) is unknown, samples with given process parameters have to be produced in the first place and then the corresponding values of the descriptor variables have to be measured. In practice, the process of determining descriptors for a set of process parameters is often expensive and associated with statistical uncertainty.
Therefore, powerful methods for multiobjective optimization on a multivariate decision space are needed. A few standard approaches are implemented as packages for the statistical software \emph{R}, such as the NSGA-II algorithm by Deb et al. (2002) in the packages \emph{nsga2R} and \emph{mco}, SMS-EMOA by Beume, Naujoks, and Emmerich (2007) in the \emph{ecr} package and MOEA/D by Zhang and Li (2007) in the package \emph{MOEADr}.
These approaches are evolutionary algorithms,
which deliver the experimenter a whole set of Pareto optimal solutions.
In Pareto optimization there is frequently the assumption that no single solution exists, which simultaneously optimizes all objectives and therefore a trade-off between conflicting objectives must be accepted.
Therefore, in this article it is assumed that target values with surrounding acceptance regions have been defined for which a solution exists, however the search space is too big for exploration by random or grid search evaluations.
In this project, the task is to find at least one solution with outcome in a previously determined acceptance region around the target value \(\mathbf{d}^*\), under the restriction that only a limited number of evaluations can be performed.
A sequential approach using dimension reduction techniques has been developed to achieve this. The idea of applying dimension reduction in sequential optimization has been pursued e.g., in Vijayakumar and Schaal (2000) and Winkel et al. (2020).
Simulations have been performed to compare the performance of the new approach and the above-mentioned algorithms. This article
focuses on finding predictor values whose corresponding descriptor values are as close as possible to the optimal value, while using only a pre-specified number of observations.
The simulation study also accounts for the statistical nature of the data, which is due to measurement errors in the descriptor variable, uncertainties in the tuning of the process parameters and randomness due to inhomogeneity in the material.
A simplified version of the proposed algorithm has been applied for single objective optimization, cf. Bader et al. (2019). The more general version presented here has been developed for multi-objective optimization problems.

\hypertarget{the-algorithm}{%
\section{The algorithm}\label{the-algorithm}}

The main steps of the proposed algorithm are depicted in Figure \ref{fig:framework}.
After an initialization process the first iteration starts with dimension reduction of the descriptor space based on principal component analysis (PCA). Then one or more sequential blocks of dimension reduction of the process parameters using partial least squares (PLS, also \emph{Projection to Latent Structure}), univariate conditional model estimation, and optimization follow. After that the back transformation of the suggested solution to the original space follows and the measurements are performed. Finally, a validation checks if the process parameter candidates can be considered as a solution. If the candidates are discarded, a new iteration begins. The process is repeated until a pre-specified number of iterations is reached.

\tikzset{every picture/.style={line width=0.75pt}}
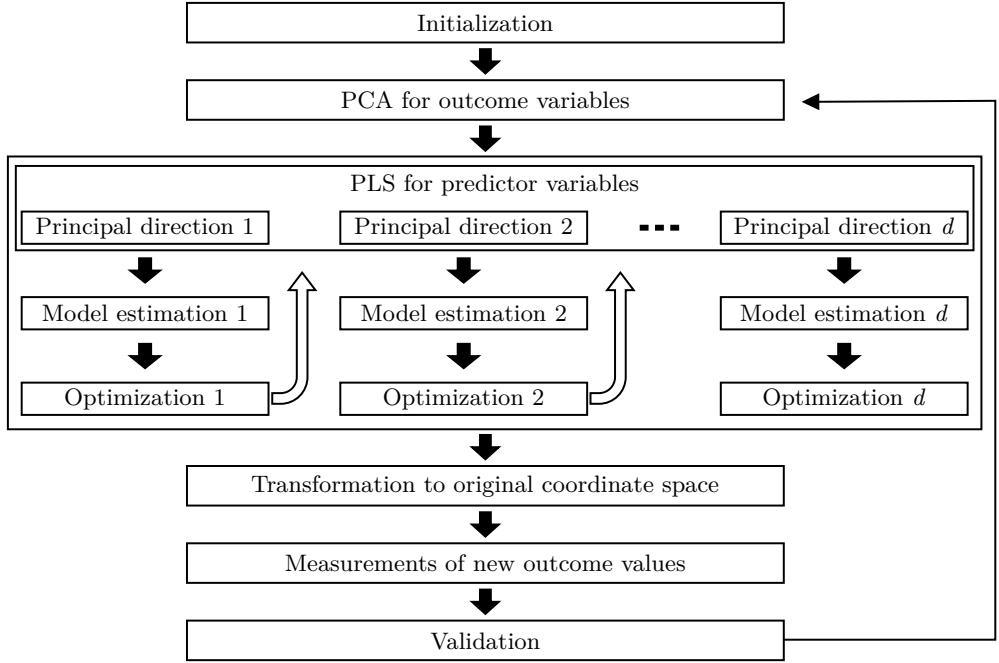
\begin{figure}[ht]
\centering

\tikzset{every picture/.style={line width=0.75pt}} 

\begin{tikzpicture}[x=0.75pt,y=0.75pt,yscale=-1,xscale=1]

\draw   (92.32,3.25) -- (393.32,3.25) -- (393.32,23.22) -- (92.32,23.22) -- cycle ;

\draw  [fill={rgb, 255:red, 0; green, 0; blue, 0 }  ,fill opacity=1 ] (236.04,32.79) -- (240.09,32.79) -- (240.09,26.6) -- (245.56,26.6) -- (245.56,32.79) -- (249.61,32.79) -- (242.82,38.88) -- cycle ;
\draw   (2,80.77) -- (493,80.77) -- (493,218.46) -- (2,218.46) -- cycle ;

\draw   (8.96,108.57) -- (133.45,108.57) -- (133.45,124.86) -- (8.96,124.86) -- cycle ;

\draw   (8.96,151.75) -- (133.45,151.75) -- (133.45,168.04) -- (8.96,168.04) -- cycle ;

\draw   (8.96,194.92) -- (133.45,194.92) -- (133.45,211.21) -- (8.96,211.21) -- cycle ;

\draw  [fill={rgb, 255:red, 0; green, 0; blue, 0 }  ,fill opacity=1 ] (63.67,138.35) -- (68.17,138.35) -- (68.17,132.16) -- (74.24,132.16) -- (74.24,138.35) -- (78.74,138.35) -- (71.2,144.44) -- cycle ;
\draw  [fill={rgb, 255:red, 0; green, 0; blue, 0 }  ,fill opacity=1 ] (63.67,181.53) -- (68.17,181.53) -- (68.17,175.34) -- (74.24,175.34) -- (74.24,181.53) -- (78.74,181.53) -- (71.2,187.62) -- cycle ;
\draw   (135.33,205.6) -- (139.62,205.6) .. controls (147.06,205.6) and (153.1,199.56) .. (153.1,192.12) -- (153.1,148.24) -- (157.14,148.24) -- (150.4,139.38) -- (143.67,148.24) -- (147.71,148.24) -- (147.71,192.12) .. controls (147.71,196.59) and (144.09,200.21) .. (139.62,200.21) -- (135.33,200.21) -- cycle ;

\draw   (169.65,108.57) -- (294.15,108.57) -- (294.15,124.86) -- (169.65,124.86) -- cycle ;

\draw   (169.65,151.75) -- (294.15,151.75) -- (294.15,168.04) -- (169.65,168.04) -- cycle ;

\draw   (169.65,194.92) -- (294.15,194.92) -- (294.15,211.21) -- (169.65,211.21) -- cycle ;

\draw   (295.83,205.6) -- (300.13,205.6) .. controls (307.57,205.6) and (313.6,199.56) .. (313.6,192.12) -- (313.6,148.24) -- (317.64,148.24) -- (310.91,139.38) -- (304.17,148.24) -- (308.22,148.24) -- (308.22,192.12) .. controls (308.22,196.59) and (304.59,200.21) .. (300.13,200.21) -- (295.83,200.21) -- cycle ;
\draw  [fill={rgb, 255:red, 0; green, 0; blue, 0 }  ,fill opacity=1 ] (224.37,138.35) -- (228.86,138.35) -- (228.86,132.16) -- (234.93,132.16) -- (234.93,138.35) -- (239.43,138.35) -- (231.9,144.44) -- cycle ;
\draw  [fill={rgb, 255:red, 0; green, 0; blue, 0 }  ,fill opacity=1 ] (224.37,181.53) -- (228.86,181.53) -- (228.86,175.34) -- (234.93,175.34) -- (234.93,181.53) -- (239.43,181.53) -- (231.9,187.62) -- cycle ;

\draw   (361.55,108.57) -- (486.04,108.57) -- (486.04,124.86) -- (361.55,124.86) -- cycle ;

\draw   (361.55,151.75) -- (486.04,151.75) -- (486.04,168.04) -- (361.55,168.04) -- cycle ;

\draw   (361.55,194.92) -- (486.04,194.92) -- (486.04,211.21) -- (361.55,211.21) -- cycle ;

\draw  [fill={rgb, 255:red, 0; green, 0; blue, 0 }  ,fill opacity=1 ] (416.26,181.53) -- (420.76,181.53) -- (420.76,175.34) -- (426.83,175.34) -- (426.83,181.53) -- (431.33,181.53) -- (423.8,187.62) -- cycle ;
\draw  [fill={rgb, 255:red, 0; green, 0; blue, 0 }  ,fill opacity=1 ] (416.26,138.35) -- (420.76,138.35) -- (420.76,132.16) -- (426.83,132.16) -- (426.83,138.35) -- (431.33,138.35) -- (423.8,144.44) -- cycle ;

\draw  [fill={rgb, 255:red, 0; green, 0; blue, 0 }  ,fill opacity=1 ] (321.15,115.73) -- (324.62,115.73) -- (324.62,117.7) -- (321.15,117.7) -- cycle ;
\draw  [fill={rgb, 255:red, 0; green, 0; blue, 0 }  ,fill opacity=1 ] (336.58,115.73) -- (340.05,115.73) -- (340.05,117.7) -- (336.58,117.7) -- cycle ;
\draw  [fill={rgb, 255:red, 0; green, 0; blue, 0 }  ,fill opacity=1 ] (328.87,115.73) -- (332.34,115.73) -- (332.34,117.7) -- (328.87,117.7) -- cycle ;

\draw   (6,85.55) -- (489,85.55) -- (489,128.12) -- (6,128.12) -- cycle ;

\draw   (92.32,42.27) -- (393.32,42.27) -- (393.32,62.24) -- (92.32,62.24) -- cycle ;

\draw   (92.32,315.03) -- (393.32,315.03) -- (393.32,335) -- (92.32,335) -- cycle ;

\draw   (92.32,276.01) -- (393.32,276.01) -- (393.32,295.98) -- (92.32,295.98) -- cycle ;

\draw   (92.32,236.99) -- (393.32,236.99) -- (393.32,256.96) -- (92.32,256.96) -- cycle ;

\draw    (393,324.67) -- (500,324.67) -- (500,52.67) -- (405.3,53.19) ;
\draw [shift={(402.3,53.2)}, rotate = 359.69] [fill={rgb, 255:red, 0; green, 0; blue, 0 }  ][line width=0.08]  [draw opacity=0] (8.93,-4.29) -- (0,0) -- (8.93,4.29) -- cycle    ;
\draw  [fill={rgb, 255:red, 0; green, 0; blue, 0 }  ,fill opacity=1 ] (236.04,71.81) -- (240.09,71.81) -- (240.09,65.62) -- (245.56,65.62) -- (245.56,71.81) -- (249.61,71.81) -- (242.82,77.9) -- cycle ;
\draw  [fill={rgb, 255:red, 0; green, 0; blue, 0 }  ,fill opacity=1 ] (236.04,305.55) -- (240.09,305.55) -- (240.09,299.37) -- (245.56,299.37) -- (245.56,305.55) -- (249.61,305.55) -- (242.82,311.65) -- cycle ;
\draw  [fill={rgb, 255:red, 0; green, 0; blue, 0 }  ,fill opacity=1 ] (236.04,266.53) -- (240.09,266.53) -- (240.09,260.35) -- (245.56,260.35) -- (245.56,266.53) -- (249.61,266.53) -- (242.82,272.63) -- cycle ;
\draw  [fill={rgb, 255:red, 0; green, 0; blue, 0 }  ,fill opacity=1 ] (236.04,227.52) -- (240.09,227.52) -- (240.09,221.33) -- (245.56,221.33) -- (245.56,227.52) -- (249.61,227.52) -- (242.82,233.61) -- cycle ;

\draw (147.95,120.18) node   [align=left] { \ \ };
\draw (308.38,109.76) node   [align=left] { \ \ };
\draw (242.82,246.98) node   [align=left] {{\footnotesize Transformation to original coordinate space}};
\draw (242.82,286) node   [align=left] {{\footnotesize Measurements of new outcome values}};
\draw (242.82,325.02) node   [align=left] {{\footnotesize Validation}};
\draw (242.82,52.25) node   [align=left] {{\footnotesize PCA for outcome variables}};
\draw (247.5,95.17) node   [align=left] {{\footnotesize PLS for predictor variables}};
\draw (247.5,106.84) node   [align=left] {};
\draw (423.8,203.06) node   [align=left] {{\footnotesize Optimization \textit{d}}};
\draw (423.8,159.89) node   [align=left] {{\footnotesize Model estimation \textit{d}}};
\draw (423.8,116.71) node   [align=left] {{\footnotesize Principal direction \textit{d}}};
\draw (231.9,203.06) node   [align=left] {{\footnotesize Optimization 2}};
\draw (231.9,159.89) node   [align=left] {{\footnotesize Model estimation 2}};
\draw (231.9,116.71) node   [align=left] {{\footnotesize Principal direction 2}};
\draw (71.2,203.06) node   [align=left] {{\footnotesize Optimization 1}};
\draw (71.2,159.89) node   [align=left] {{\footnotesize Model estimation 1}};
\draw (71.2,116.71) node   [align=left] {{\footnotesize Principal direction 1}};
\draw (247.5,149.62) node   [align=left] {};
\draw (242.82,13.23) node   [align=left] {{\footnotesize Initialization}};
\end{tikzpicture}
\vspace*{0pt}\caption{Framework of the suggested algorithm.} \label{fig:framework}
\end{figure}

In the initialization step basic parameters for the algorithm are set:
The target value \(\mathbf{d}^*\) for the optimization has already been mentioned above. Additionally, a target region is defined, which includes all descriptor variable combinations which are considered as acceptable solution. Typically, this would be defined as a rectangular region with the target value in the centre, which is determined by the minimally and maximally acceptable values for each descriptor outcome, \(d_i^{(min)}=d_i^*-\delta_i\), \(d_i^{(max)}=d_i^*+\delta_i\), for \(\delta_i > 0\), \(i=1,\dots,D\).
To limit the search space, minimum and maximum values for each predictor variable \(p_i^{(min)}\), \(p_i^{(max)}\), \(i=1,...,P\) are set.
Furthermore, the maximum number of iterations \(r\) has to be set, i.e., after \(r\) iterations of the algorithm and subsequently new measurements the algorithm will terminate unsuccessfully, if no solution has been found.
An optional parameter is the smallest distinguishable unit \(\Delta_1,\dots \Delta_P\) for each predictor. New suggested parameters which are in the scope of the experiment indistinguishably close to previous measurements will be avoided.
Another optional parameter, \(d\), fixes the number of principal components that will be used for the dimension reduction of the predictors. This has to be a positive integer smaller or equal to the dimension of the predictor space.
If no data sample is given in advance, it has to be created in the initialization step: a set of a small number \(k\) of supporting points has to be chosen and a number \(l\) of independent measurements of the experiments have to be performed at each of the supporting points, thus creating a sample of premeasurements with size \(k \cdot l\).
To decide for the initial supporting points it is suggested to follow general good practice in experimental design. In the special case where the observable variance for the descriptor variables is zero, one would obviously set \(l=1\), i.e., measurements are not repeated. Otherwise, the decision for a specific choice of \(l\) will be driven by the size of the variance on the one hand and the costs for additional repeated measurements on the other hand. To ensure samples that are well spread over the parameter space, low discrepancy sampling methods should be preferred to random sampling. A good starting point for such methods is the popular Latin hypercube design, cf. McKay, Beckman, and Conover (1979). Further examples are Sobol sequences, Sobol (1967), orthogonal arrays, Owen (1992) and further variations of Latin hypercubes as e.g., Tang (1993), Ye, Li, and Sudjianto (2000), Husslage et al. (2010).
In the following the initial measurements will be considered as given in advance, i.e., a data set \((\mathbf p^{(i)}, \mathbf d^{(i)})\), \(i=1,\dots,n\),
\(n:=k\cdot l\), is assumed.

\begin{figure}[h]
\centering
\vspace*{20pt}

\begin{center}\includegraphics[width=1\linewidth]{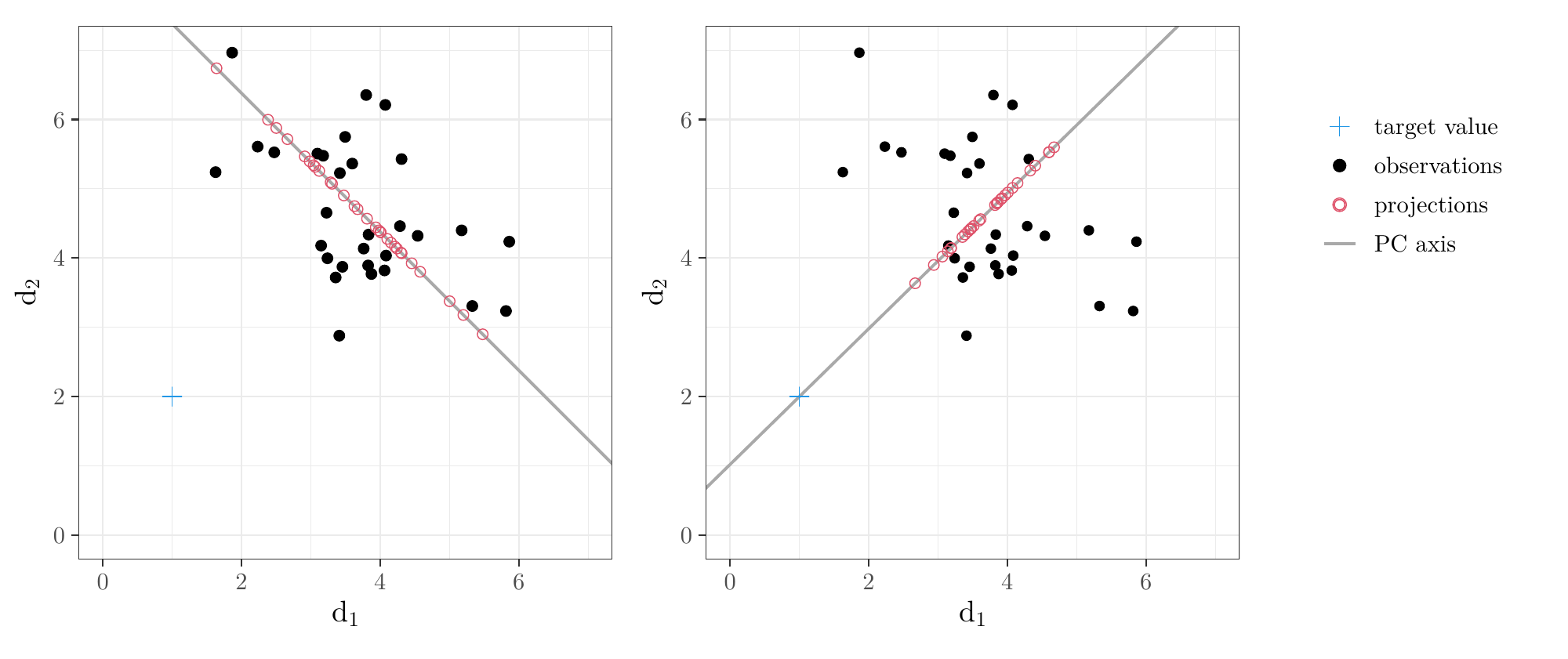} \end{center}

\vspace*{-20pt}\caption{Traditional choice of the principal component (PC) axis in PCA (left) vs. forcing a specific value (here: target value) to lie exactly on the principal component axis (right).} \label{fig:pcaprojection}
\end{figure}

In the following the problem of finding optimal process parameters, i.e., finding \(\mathbf{p}\) such that
\(\phi(\mathbf{p}):=E[\mathbf{d}|\mathbf{p}]=\mathbf{d}^*,\)
is simplified to finding solutions for the lower dimensional relationship
\(f(\mathbf{x}):=E[y|\mathbf{x}]=0\)
with \emph{pseudo predictor} variable \(\mathbf{x}=(x_1,\dots,x_d)\), \(1\leq d\leq P\) and the one-dimensional \emph{pseudo descriptor} \(y \in \mathbb{R}\).
At first the descriptor space is taken care of:
If the descriptor space has dimension \(D>1\), PCA will be performed on the descriptor variables.
Other than in most common applications of PCA, here the data is not centred to have zero mean, but rather the descriptor target value \(\mathbf{d}^*\) is chosen as the new centre of the data. Scaling to variance one is performed as usual.
Figure \ref{fig:pcaprojection} depicts the resulting first principal component axis for a random bivariate sample data set for both, using traditional standardization (left) and shifting the target value to the origin (right).

If the data are centred before applying the PCA (left plot in Figure \ref{fig:pcaprojection}), the mean vector becomes the origin and the first principal component points in the direction where the data have the largest possible variance.
If, however, the data are shifted in such a way that the target value becomes the new origin, the first principal component maximizes the sum-of-squared deviations from the target value instead.
It is proposed that this approach is more appropriate for the following modelling step.
The descriptor data which are transformed in this way are denoted as row vectors
\(\mathbf{z}^{(i)}=\left(\frac{d_1^{(i)}-d_1^*}{\hat \sigma_1} ,\cdots,\frac{d_{D}^{(i)}-d_D^*}{\hat \sigma_D}\right),\)
where \(\hat \sigma_j\) is the empirical standard deviation of the observations \(d^{(i)}_j\), \(j=1,\dots,D\) for each \(i=1,\dots,n\).
As a result of the PCA the loadings matrix \(\mathbf{W}\) and the component scores \(\mathbf{y}^{(i)}\), \(i=1,\dots,D\), are determined.
The loadings matrix \(\mathbf{W}\) is a \(D \times D\) matrix which has the so-called principal components as its columns.
The component score \(\mathbf{y}^{(i)}\) is the vector of the components of \(\mathbf{z}^{(i)}\) with respect to the base vectors given in the columns of \(\mathbf{W}\).
For the principal component decomposition the relationship
\({\mathbf{y}^{(i)}}^T= \mathbf{z}^{(i)} \mathbf{W}\) holds.
In the following only the first principal component will be considered, leaving us with a one-dimensional descriptor space. The one-dimensional pseudo descriptor will be identified as \(y_i := {y}^{(i)}_{1}\). To avoid unnecessary distinction of the two cases \(D=1\) and \(D>1\) in the following, the above-described transformation is also performed in case of \(D=1\), i.e., \(y_i := {z}^{(i)}_{1}\).

In a next step the PLS1-algorithm of Wold (1966) is used to determine \(d\) principal components with maximal correlation to the pseudo-descriptor \(y_i\).
This leaves us with pseudo-data \((\mathbf{x}^{(i)}, y_i)\) from a \(d \times 1\)-relationship
\(f: \mathbf{x} \mapsto E[y|\mathbf{x}], \, \mathbb{R}^d \to \mathbb{R}.\)
\(d\) might be fixed or chosen by a data-based approach, e.g., the Kaiser-Guttman criterion (Guttman 1954), which is widely known as \emph{Kaiser's rule}.
If no other choice for the parameter \(d\) has been set in the initialization step, the algorithm will simply choose \(d=D\), i.e., the dimension of the process parameters is not reduced.

The first model estimation step starts with modelling the relationship
\(f_1: x_1 \mapsto E[y|x_1], \, \mathbb{R} \to \mathbb{R},\)
as polynomial regression model, i.e., the stochastic root finding problem is replaced by a surrogate deterministic root finding problem. As regression models polynomials of orders \(m=1\) to \(5\) are permitted. The actual order \(m\) is chosen based on the BIC (Bayesian information criterion), i.e.,
\(\operatorname{BIC}=\ln(n)k-2 \ln(\hat L),\)
where \(\hat L\) is the maximized value of the likelihood function of the fitted polynomial model, \(n\) is the number of observations and \(k\) the number of parameters estimated by the model.
The model with the lowest BIC is preferred.
As a result, a model estimation \(\hat f_1(x_1)=\sum\limits_{i=0}^m \alpha_i x_1^i\) with regression coefficients \(\alpha_i\), \(i=0,\dots,m\) is determined.

In the following optimization step possible solutions, i.e., vectors of predictor values \(\mathbf{{\hat p}}\) which generate the desired vector of target descriptor values \(\mathbf{d}^*\) according to the one-dimensional model \(y=f_1(x_1)+\varepsilon\), have to be found.
Candidates for appropriate pseudo predictors will be determined as the roots of
\(\hat f_1(x_1) = \sum_{i=0}^m \alpha_i ({x_1})^i.\)
Let \({\hat x}_1\) denote one of the estimated roots.
Denote the predictor variables of the \(n\) so far available measurements jointly as matrix \(\mathbf{P}=\left(\mathbf{p}^{(1)},\dots,\mathbf{p}^{(n)}\right)^T\). The above application of the PLS1-algorithm to the standardized predictor variables returns corresponding scores (pseudo predictor variables), which are now denoted as matrix \(\mathbf{X}=\left(\mathbf{x}^{(1)},\dots,\mathbf{x}^{(n)}\right)^T\), together with the modified weights matrix \({\mathbf{V}}\), which satisfies
\(\mathbf{P}=\mathbf{X}{\mathbf{V}}.\)
Denote by \({\mathbf{V}}_{1,\cdot}\) the first row of \({\mathbf{V}}\).
Then, the coordinates of the pseudo predictor \({\hat x}_1\) in the standardized predictor space can be determined as \(\mathbf{z}\),
\(\mathbf{z}^T := {\hat x}_1\cdot {\mathbf{V}}_{1,\cdot}\)

The coordinates in the original predictor space are then \(\mathbf{\hat p}=({\hat p}_1,\dots, {\hat p}_{P})^T\), with \({\hat p}_i = z_i\cdot \hat \sigma_i+\hat \mu_i\), where \(\hat \sigma_i\) and \(\hat \mu_i\) are the empirical variance and mean of the observed predictors.
Under the simplifying assumption that the relationship \(f: \mathbf{x} \mapsto E[y|\mathbf{x}]\) is mainly driven by the effect of the first principal component, i.e., that \(E[y|\mathbf{x}] \approx E[y|x_1]\),
\(\mathbf{\hat p}\) can be suggested as reasonable approximation to a pseudo predictor
which corresponds to a pseudo descriptor close to zero. The next steps of the sequential approach will first be outlined for the second parameter value \(x_2\), before the general procedure is stated in more detail.

To improve the candidate for the optimization problem, it is assumed next that \(E[y|\mathbf{x}] \approx E[y|x_1,x_2]\) is an appropriate approximation and thus \(y=f_2(x_1, x_2)+\varepsilon\) with \(E(\varepsilon|x_1,x_2)=0\).
In the following consider the coordinate for the first principal component \(x_1\) as given; for sake of simplicity, \(x_1:= {\hat x}_1\) is chosen, as determined in the process above. If an \({\hat x}_{2}\) can be found such that \(f_2({\hat x}_1, {\hat x}_{2})=0\), then the pair of pseudo predictor values \(({\hat x}_1, {\hat x}_{2})\) is a proper candidate for a solution in the bivariate space of the first two principal components.

\begin{figure}[h]

\begin{center}\includegraphics[width=0.65\linewidth]{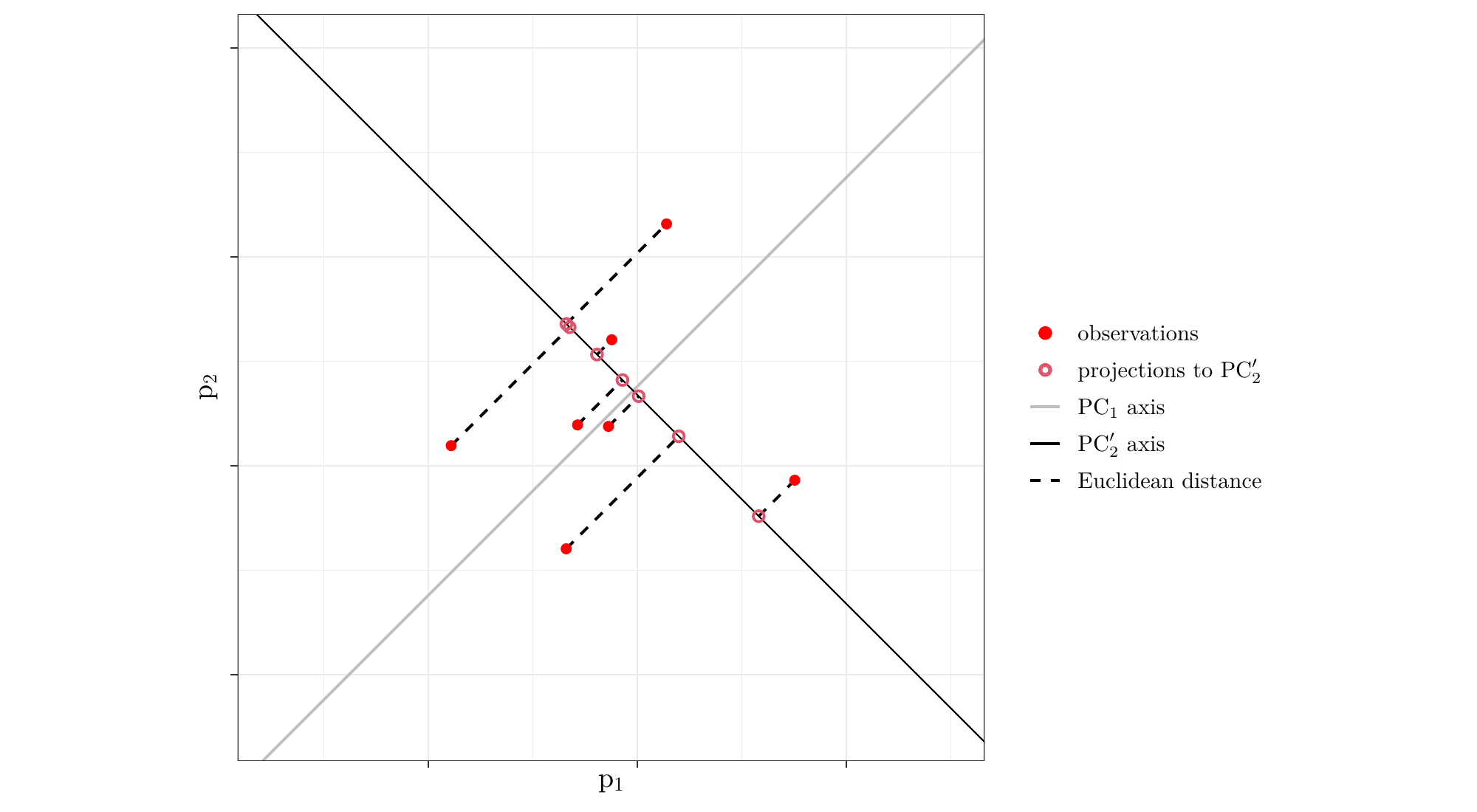} \end{center}
\caption{Euclidean distances between observations $x_1^{(1)},\dots,x_1^{(n)}$ and their projections.} \label{fig:eucdist}
\end{figure}

To find such a solution, a model for the function \(x_2\to y=f({\hat x}_1,x_2)\) is proposed.
A priori there would usually be no observations with the first pseudo predictor coordinate being equal to the specified coordinate value \({\hat x}_1\). To create a conditional model based on the available data, a weighted regression approach will be applied: A weighted regression model for \(y \sim x_2\) given \(x_1= {\hat x}_1\) shall serve as an estimate for the marginal distribution in the direction of the second principal component axis.
The sample data is mainly weighted by the inverse of the squared Euclidean distance of the observations \(x_1^{(1)},\dots,x_1^{(n)}\) to the line \(PC'_2\), the second principal component shifted from the origin along the first principal component by \({\hat x}_1\) units, such that the line passes through \(\hat x_1\), cf.~Figure \ref{fig:eucdist}.

Now the general approach will be outlined. If values for \(x_1,\dots,x_{j-1}\) are given, assume for the next sequential modelling step \(E[y|\mathbf{x}] \approx E[y|x_1,\dots,x_j]\), i.e.,
\(y=E[y|x_1,\dots, x_j]+\varepsilon=f_j(x_1,\dots, x_j)+\varepsilon,\)
with error term \(\varepsilon\), \(E[\varepsilon|x_1,\dots,x_j]=0\).
Consider now \(x_1,\dots,x_{j-1}\) as fixed, where \(x_1:= {\hat x}_1,\dots,x_{j-1}:= {\hat x}_{j-1}\).
If there is an \({\hat x}_{j}\) such that \(f_j({\hat x}_1,\dots, {\hat x}_{j})=0\), then \(({\hat x}_{1},\dots,{\hat x}_{j})\) can be considered as an appropriate candidate for a pseudo predictor variable in the space of the first \(j\) principal components. Based on the conditional model
\(y=f_{j}(x_1={\hat x}_1,\dots,x_{j-1}={\hat x}_{j-1}, x_j )+\varepsilon,\)
\(f_{j}({\hat x}_1,\dots,{\hat x}_{j-1}, \cdot)\) will be modelled as polynomial function of order \(1\leq m\leq 5\) using weighted least squares (WLS) regression.
The weights will be chosen as the inverse of the squared Euclidean distance between the so far evaluated pseudo predictor coordinates \((x^{(i)}_1,\dots,x^{(i)}_{j-1})\), \(i=1,\dots,n\) and their projections onto \(PC_j'\), the \(j\)-th principal component, shifted from the origin along the \(j-1\) first principal components by \({\hat x}_1,\dots,{\hat x}_{j-1}\), respectively.
Coefficients for the polynomial will be estimated by
\[\operatorname{WLS}(\boldsymbol \beta)= \underset{\boldsymbol \beta}{\operatorname{arg\,min}} \sum_{i=1}^n \frac{1}{w_i}(y_i-\mathbf x{^{(i)}}^{\top} \boldsymbol \beta)^2,\]
where the weight function is chosen as
\[w_i=\sum \limits_{k\in\{1,\dots,j-1\}} \left({x^{(i)}_k-{\hat x}_{k}}\right)^2 + \sum \limits_{l\in\{2,\dots,p\}} {y^{(i)}_l}^2.
\label{eq:weights}
\tag{1}\]
Recall that the weighted regression is based on the first principal component in the pseudo descriptor space only. The second term of equation \ref{eq:weights} was added for models with more than one response, as it is proposed that the weight for the \(i\)-th observation
should also consider how close the pseudo-observation \(\mathbf{y}^{(i)}\) is to the chosen principal axis of the pseudo descriptor space.

The newly estimated model is now \(\hat f_{j}(x_j)=\sum\limits_{k=0}^m \beta^{(j)}_k x_j^k\) with coefficients \(\beta^{(j)}_k\), \(k=1,\dots,m\).
Analogous to the above, solutions in the original predictor space are determined:
At first estimate \({\hat x}_{j}\) as a root of \(\hat f_{j}(\cdot)\).
Then transform \(({\hat x}_1, \dots, {\hat x}_{j})\) to the standardized predictor space by multiplication with the corresponding rows of \({\mathbf{V}}\), i.e.,
\(({\hat x}_1, \dots,{\hat x}_{j})\cdot{\mathbf{V}}_{(1,\dots,j),\cdot}\).
Finally, \emph{destandardization} provides the coordinates in the original parameter space.

Unfortunately, it cannot be guaranteed that any solutions will be found, as the specified model equation might not have any roots at all.
If for some \(j\in\{1,\dots,d\}\) the equation
\(\hat f_{j}(x_j)=\sum\limits_{k=0}^m \beta^{(j)}_k x_j^k=0\)
cannot be solved by any \(x_j\), using this method, no solution can be found with the currently available data.
As long as the maximal number of iterations \(r\) is not reached, it is assumed that a solution exists and new data points are searched in order to proceed with further iterations of the algorithm. If no expert knowledge is available to find further regions that are likely informative for the modelling process of the relationship between predictors and descriptors, it is advised to continue the search in not yet explored parameter regions.

Let \(\mathcal I_i\) denote the set of pseudo predictor coordinates \(x_i\) on the \(i\)-th principal component axis for which its transformation to the original predictor space \(({\hat x}_1,\dots,{\hat x}_{i-1},x_i)\cdot \mathbf{V}_{(1,\dots,i),\cdot}\) is within the initially specified boundaries, \(\prod\limits_{j=1}^{i}[p_j^{(min)},p_j^{(max)}]\).
As no solution for the \(i\)-th regression model can be found, it is suggested as a provisional fallback to set \(x_i\) to a value which is located far from the already observed coordinate values \(x_i^{(j)}\). More specifically, \({\hat x}_i\) is set in the current iteration to
\(\underset{x_i\in\mathcal I_i}{\operatorname{arg\,max}} \min_{x_i^{(j)}, \,j=1,\dots,n}|x_i-x_i^{(j)}|,\)
i.e., the \(i\)-th pseudo predictor is chosen such that its distance to the nearest observed coordinate values \(x_i^{(j)}\), \(j=1,\dots,n\) is maximized.
This currently implemented fallback works for arbitrary dimensions deterministically. It follows the vague idea that, assuming the current \(i\)-th principal direction is relevant for solving the model, filling up the so far unobserved space along its principal axis with new observations, will help to improve the models in the subsequent iterations of the algorithm.
It might be improved by adding a stochastic component to further explore the search space, but for simplicity this was not considered.

When all coordinates of the pseudo predictor \({\hat x}_1, \dots, {\hat x}_d\) have been determined and transformed to coordinates in the original process parameter space \({\hat p}_1,\dots,{\hat p}_P\), the \(l\) repeated measurements of the corresponding descriptor variables \(\hat d=\phi({\hat p}_1,\dots,{\hat p}_P)\) will be performed. These measurements will be available in subsequent iterations of the algorithm as new observations \((\mathbf p^{(i)}, \mathbf d^{(i)})\), \(i\in \{n+1,\dots,n+l\}\), where \(n\) is the sample size at the beginning of the current iteration. With the enlarged data set a new iteration can be started until the maximum number of iterations \(r\) has been reached.

When predictor values for a new measurement are determined in the optimization step, it may happen, that the new point \((p^{\text{new}}_1,\dots,p^{\text{new}}_P)\) is in close proximity to an already observed point \((p^{(j)}_1,\dots,p^{(j)}_P)\).
If it happens that \(|p^{\text{new}}_i-p^{(j)}_i|<\Delta_i\) for all \(i\in\{1,\dots,P\}\) for an observation \(j\), both predictor coordinates would be considered as practically identical.
In practice it is desirable to avoid unnecessarily performing a fixed number of iterations and allow early stopping, as soon as an appropriate solution has been found. In a deterministic environment without variance in the measurements it is a quite trivial task to check, if \(\phi({\hat p}_1,\dots,{\hat p}_P)\) lies in the target area. However, assuming the \(d_i^{(1)},\dots,d_i^{(n_i)}\) are subject to measuring errors \(\varepsilon_i\) with unknown variance, this becomes more involved. The decision, if the measured values \(\phi({\hat p}_1,\dots,{\hat p}_P)\) lie in the target area, could be done by a statistical test with a fixed significance level at each iteration of the algorithm. However, with growing number of iterations, a decision to accept the measurement as solution in the target area may become overly optimistic because of the repeated testing. Therefore, a more involved decision may be based on methods for sequential hypotheses testing under the additional consideration of the multiplicity due to the many process parameter points that are considered during the process. Bartroff and Song (2014) present a general approach on how to combine sequential tests and multiple testing. A stopping rule extending these ideas to the discussed setting will be presented in the future.

\hypertarget{simulation-study}{%
\section{Simulation study}\label{simulation-study}}

The presented algorithm has been implemented in an \emph{R}-package, cf. Mändle (2020),
which allows additional modifications. For detailed information please refer to the documentation of the package. In this article a choice of 5 modifications are considered which are denoted as approaches 1 to 5. Simulations will be performed to evaluate the performance of these approaches for several single response and multi-response models. For models with more than one response variable also the algorithms NSGA-II, SMS-EMOA and MOEA-D were applied to compare their performance to the presented approaches.

\begin{figure}[t]

\begin{center}\includegraphics[width=0.99\linewidth]{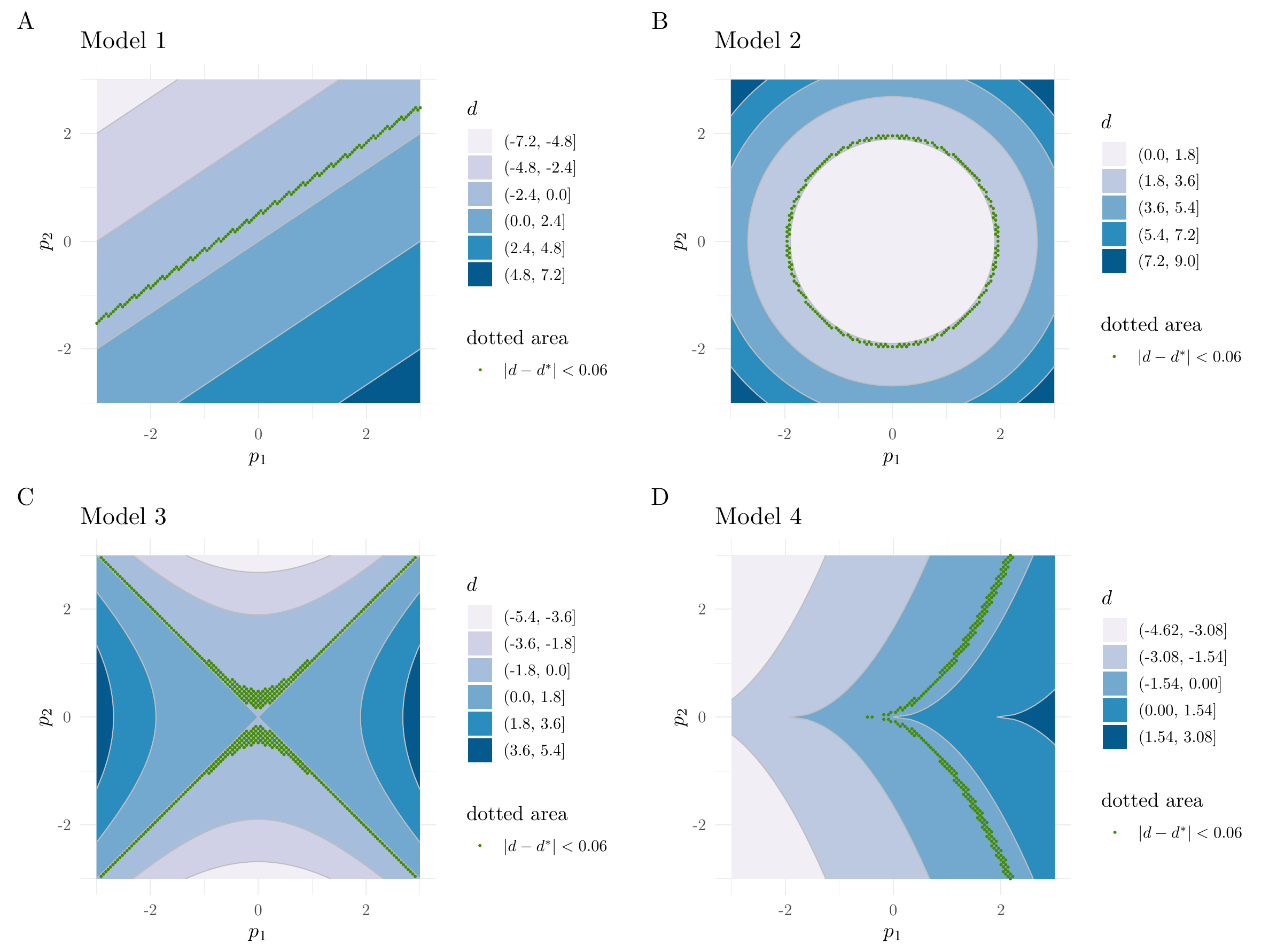} \end{center}
\caption{Distance to target value for models 1 -- 4.} \label{fig:models1234}
\end{figure}

\emph{Approach 1} uses only the first
principal component of the PLS1-transformed predictor space observations as pseudo predictors \(x^{(i)}_1\), which is basically the presented approach with \(d=1\).
\emph{Approach 2} also reduces the predictor space to dimension \(d=1\). However, a weighted regression with the weights as defined in equation \eqref{eq:weights} is performed in the prediction step. This is different from how the approach was suggested above, where weighted regression was only used for the conditional regression models when \(d>1\), whereas the unconditional models were modelled with unweighted polynomial regression.
\emph{Approach 3} is the suggested approach with \(d=2\), i.e., all principal components are being used for the prediction of the bivariate models. No actual dimension reduction of the process parameter space is performed in the bivariate models, just a transformation of the coordinates.
\emph{Approach 4} proceeds similarly as Approach 3. However, in the modelling step only those observations are used, for which the observed corresponding standardized predictor values are among the \(k=15\) unique standardized predictor values with least Euclidean distance to the measurement of the previous iteration. The idea behind this modification is to prevent the algorithm from being stuck and to lead to a faster exploration of the process parameter space.
\emph{Approach 5} is an additional approach proposed for the use in multi-objective optimization, i.e., \(D>1\). The approach is based on Approach 3; however, it is modified such that only the measurements of the last 5 evaluation points are used for the choice of the principal direction in the PCA for the descriptors. This modification aims at making the algorithm adapt faster to the new measurements.
In case that more than one solution in the search space has been found in the optimization step for a component \(\hat x_j\) of the pseudo predictor, new evaluation points will be added for each of the solutions in all approaches.

Within the simulation study for the single objective optimization, the number of iterations \(r\) is fixed to 40, i.e., always \(r=40\) iterations will be performed, even if a sufficiently good approximation to the solution has been found before, i.e., no stopping rule is applied. No smallest distinguishable unit for the parameter space is set, i.e., the only limit is the distinguishability by the machine epsilon due to rounding in floating point arithmetic.
The limit for the values of the parameter space is set to \([-5,5]^P\). The maximum order for the polynomial regression is set to 5 for all the approaches.
Each of the following simulations starts with measurements at 4 given parameter value sets of dimension \(P\). The initial evaluation points have been chosen randomly in \([-4,4]^2\), however for the simulations of each of the approaches the same sequence of random initial evaluation points has been chosen. The case without standard deviation as well as the case with a standard deviation of 0.2 in each dimension of the descriptor variable is considered. In the case with standard deviation 0.2 it is additionally observed how the performance is improved if 5 repeated measurements per evaluation point are performed.
Figure \ref{fig:models1234} shows four models for a \(2\)-dimensional predictor space and a \(1\)-dimensional descriptor space. These models have been used as true models in the following simulations of the approaches 1 -- 4 for single-response optimization. The dotted areas highlight process parameter combinations which have their corresponding descriptor value close to the target value.
Model 1 is defined as \(d=0.8\cdot p_1 - 1.2\cdot p_2 + \varepsilon\). The error term \(\varepsilon\) here and in the following models represents a normally distributed random error \(\varepsilon\) with mean zero and standard deviation \(\sigma \geq 0\).
The second and third model are represented by \(d=\alpha\cdot(p_1)^2 + \beta\cdot(p_2)^2 + \varepsilon\), where for model 2 it holds \(\alpha=\beta=0.5\) and for model 3 \(\alpha=0.5\), \(\beta=-0.5\).
The last single-response model, model 4, is defined by \(d=0.8\cdot p_1 - 1.2\cdot \sqrt{|p_2|} + \varepsilon\).

\begin{figure}[t]

\begin{center}\includegraphics[width=0.99\linewidth]{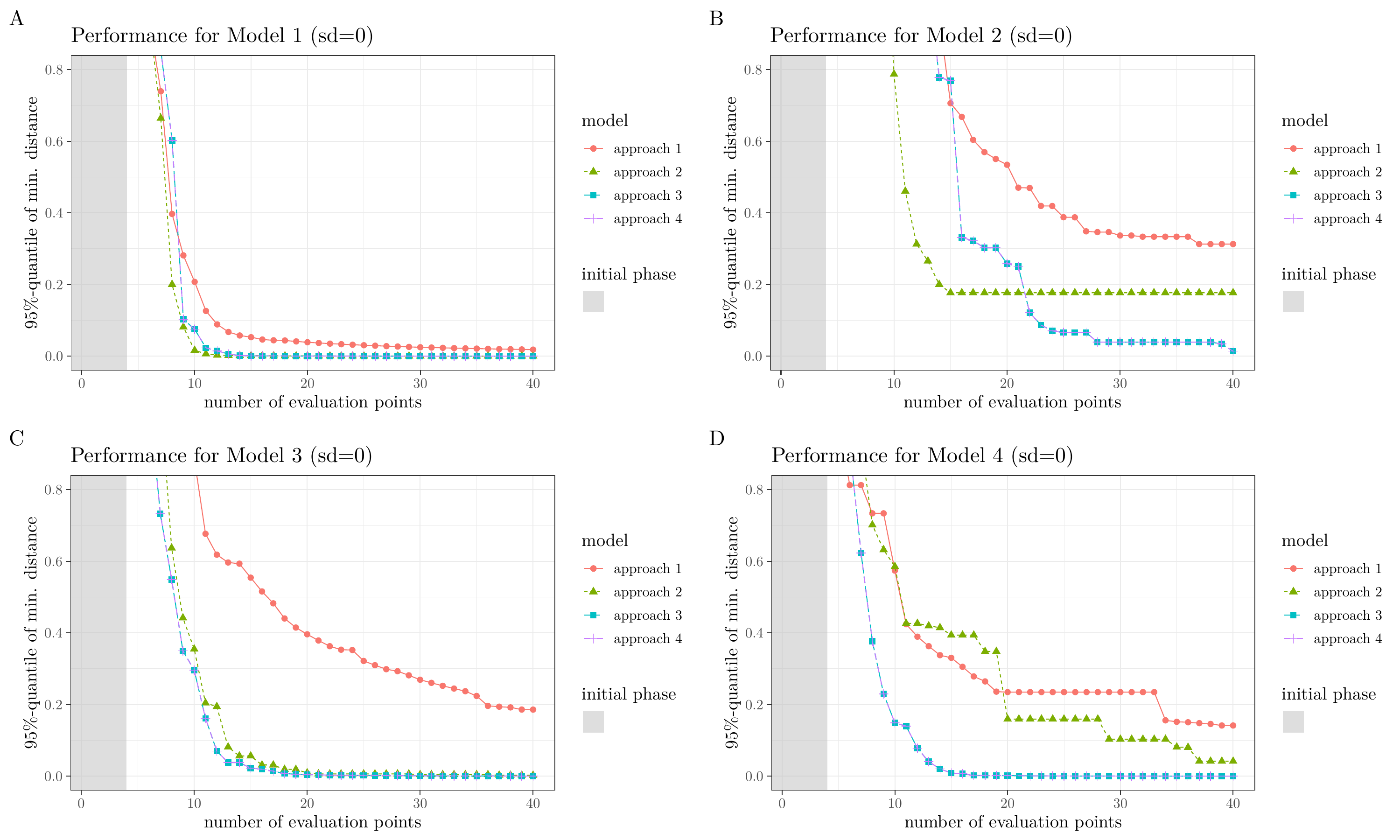} \end{center}
\caption{Performances of the approaches 1 -- 4 for models 1 -- 4 in case of exact measurements without standard deviation. Graphs for approach 3 and 4 are identical.} \label{fig:mod1234sd0performance}
\end{figure}

The results are summarized in plots A--D of figure \ref{fig:mod1234sd0performance}, where the 95\%-quantiles over 100 simulation paths of the minimum distance to the target value within the first \(i\) iterations are depicted. The results for approaches 1 to 4 are plotted as separate lines in a common coordinate system. The figure contains plots for each of the considered univariate models with \(\sigma=0\), to illustrate the performance in the case of exact measurements.
When there is no variance, the approaches work reasonably well for the considered univariate models.

\begin{figure}[ht]

\begin{center}\includegraphics[width=0.99\linewidth]{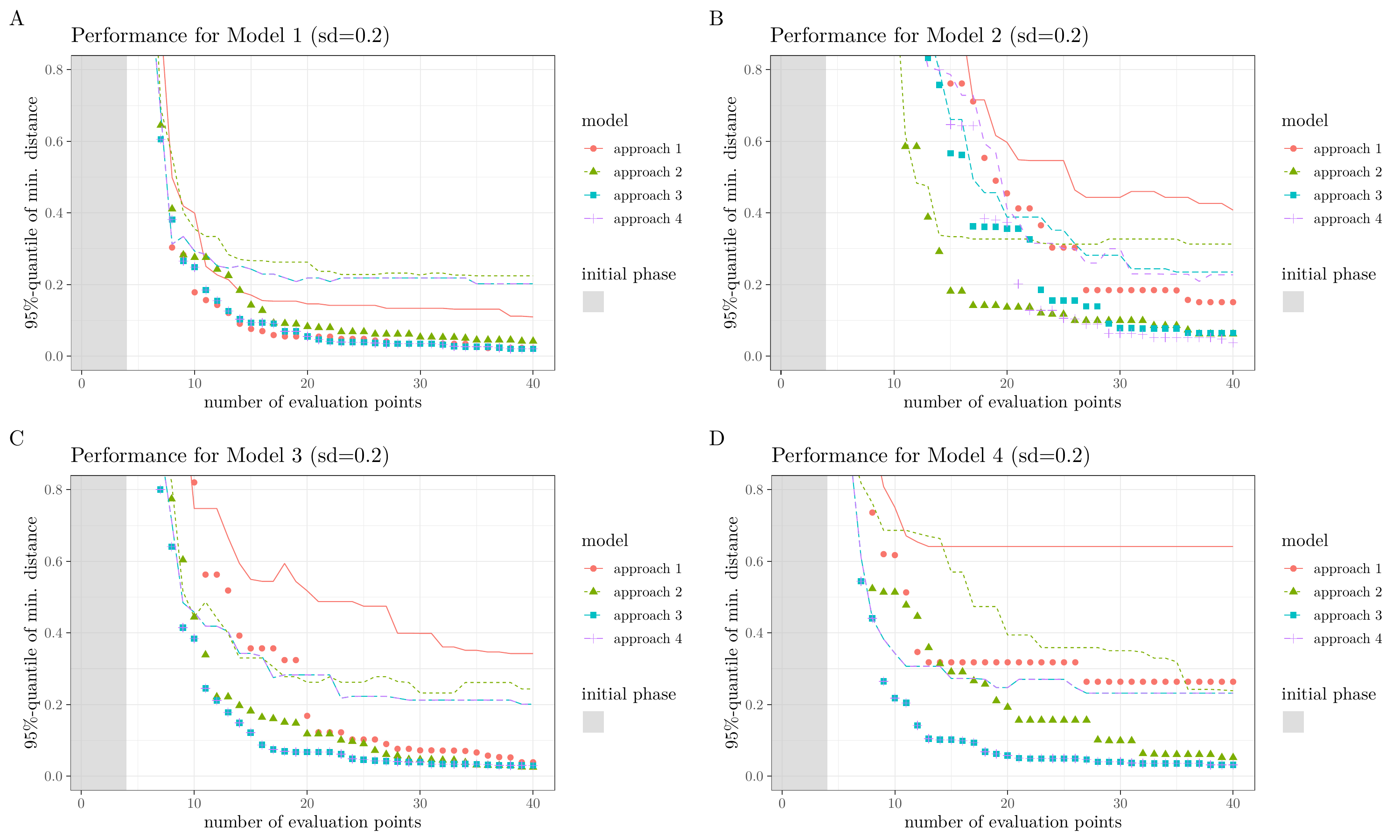} \end{center}
\caption{Performances of the approaches 1 -- 4 for models 1 -- 4 in case of standard deviation 0.2. Graphs for approach 3 and 4 are identical.} \label{fig:mod1234sd02performance}
\end{figure}
\begin{figure}[t!b!]

\begin{center}\includegraphics[width=0.99\linewidth]{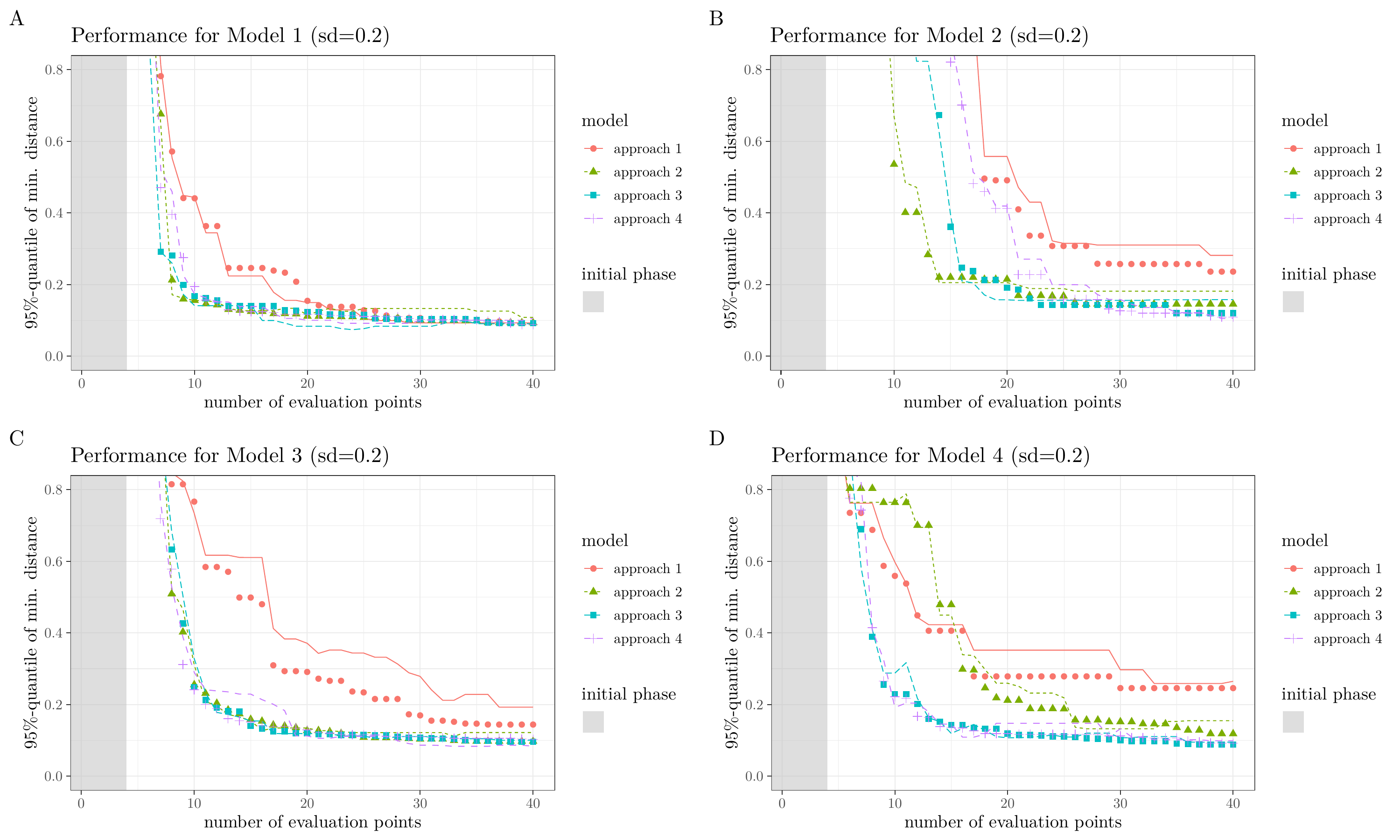} \end{center}
\caption{Performances of the approaches 1 -- 4 for models 1 -- 4 in case of standard deviation 0.2 with 5 repeated measurements per evaluation point.} \label{fig:mod1234sd02rp5performance}
\end{figure}

To investigate the effect of statistical errors in the measurements, similar plots are provided for the case of a standard deviation of 0.2, cf.~figure \ref{fig:mod1234sd02performance}. The lines, as before, depict the 95\%-quantiles of the actual, usually unknown, minimal distance to the target value. The corresponding symbols locate the corresponding observed distances to the target value, which deviate from the true difference because of the random measurement error in the target value.
As can be seen, neither of the approaches gives satisfying results. The actual distance to the target value is mostly higher than the observed distance, indicating a bias in the observed distances. Furthermore, with increasing number of evaluations it occurs that due to \emph{lucky} measurements earlier suggested approximate solutions of the optimization problem are replaced by inferior suggestions.
The plots in figure \ref{fig:mod1234sd02rp5performance} have been generated for the case where the standard deviation is still 0.2, but for every evaluation point 5 repeated measurements are evaluated. Each of the repeated measurements is considered as its own observation by the proposed algorithms. The number of evaluation points in figure \ref{fig:mod1234sd02rp5performance} therefore corresponds to 5 times the number of total measurements.
The evaluation of repeated measurements improves the performance and reduces the bias. Compared to figure \ref{fig:mod1234sd02performance} the tendency of the observed distance to the target value to be smaller than the real distance is diminished in figure \ref{fig:mod1234sd02rp5performance}. The observed distance is often even greater than the real distance to the target value. Among the so far observed model settings algorithm 4 is suggested as a good overall choice. However, a reasonable number of repeated measurements is needed to account for variance in the measurements.

\FloatBarrier

\begin{figure}[t!]

\includegraphics[width=0.99\linewidth]{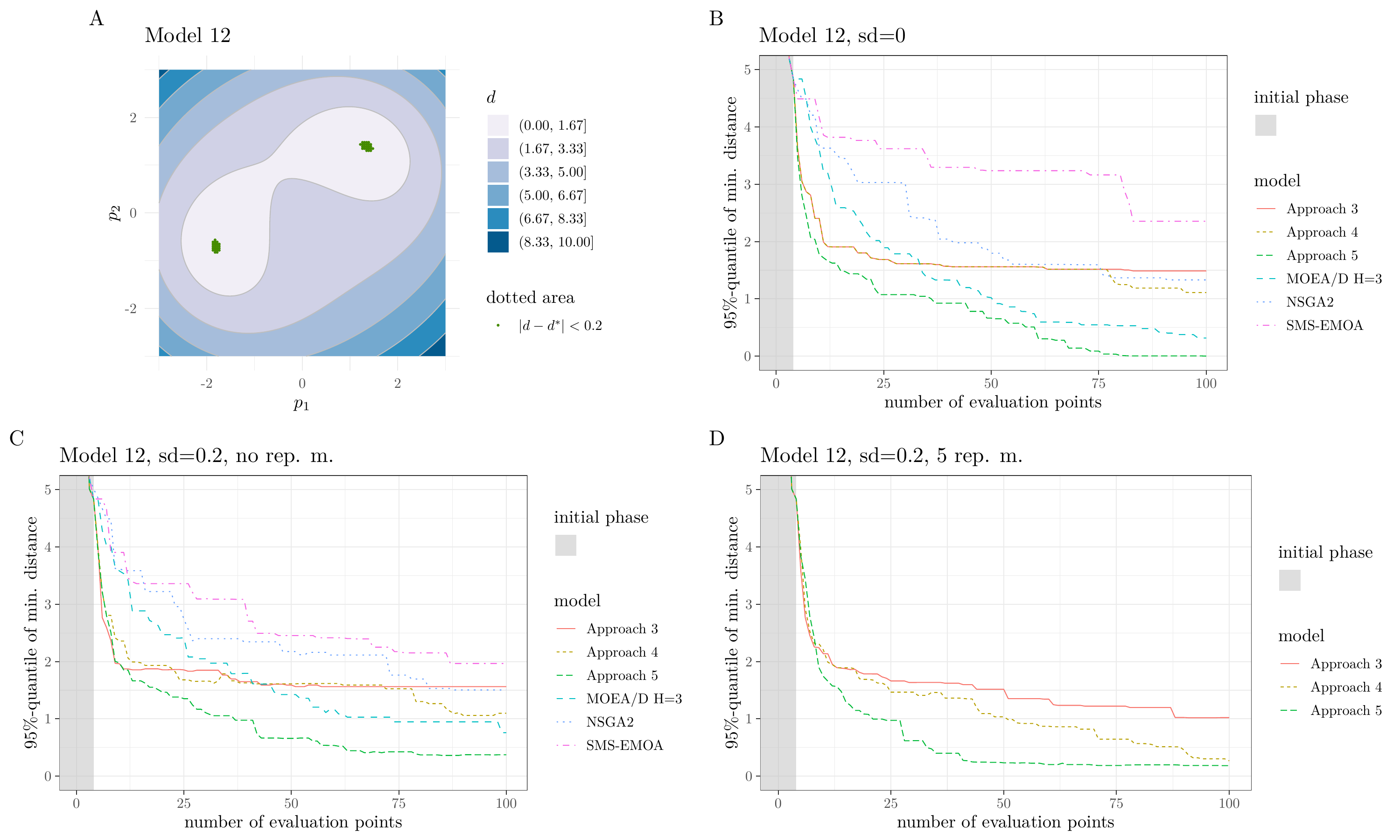} 
\caption{Performances of the approaches 3 -- 5, MOEA/D, NSGA2 and SMS-EMOA for model 12 in case of exact measurements without and with standard deviation.} \label{fig:mod12compare}
\end{figure}

Next, an example based on simulations of a bivariate 2-response model is given. In the following the number of iterations \(r\) is set to 100.
The model 12 is defined as the joint 2-response model \(\phi: \mathbf{p} \mapsto \mathbf{d}, \, \mathbb R^2 \to \mathbb R^2\) with its components given by model 1 and model 2. This model contains two target regions, each containing an exact solution.
As there are now two dimensions in the descriptor space, the model plot in figure \ref{fig:mod12compare} depicts the distance to the target \(d^*\), \(\|\phi(\mathbf{p})-\mathbf{d}^*\|_2\), instead of the target coordinates.

\begin{figure}[t!]

\includegraphics[width=0.99\linewidth]{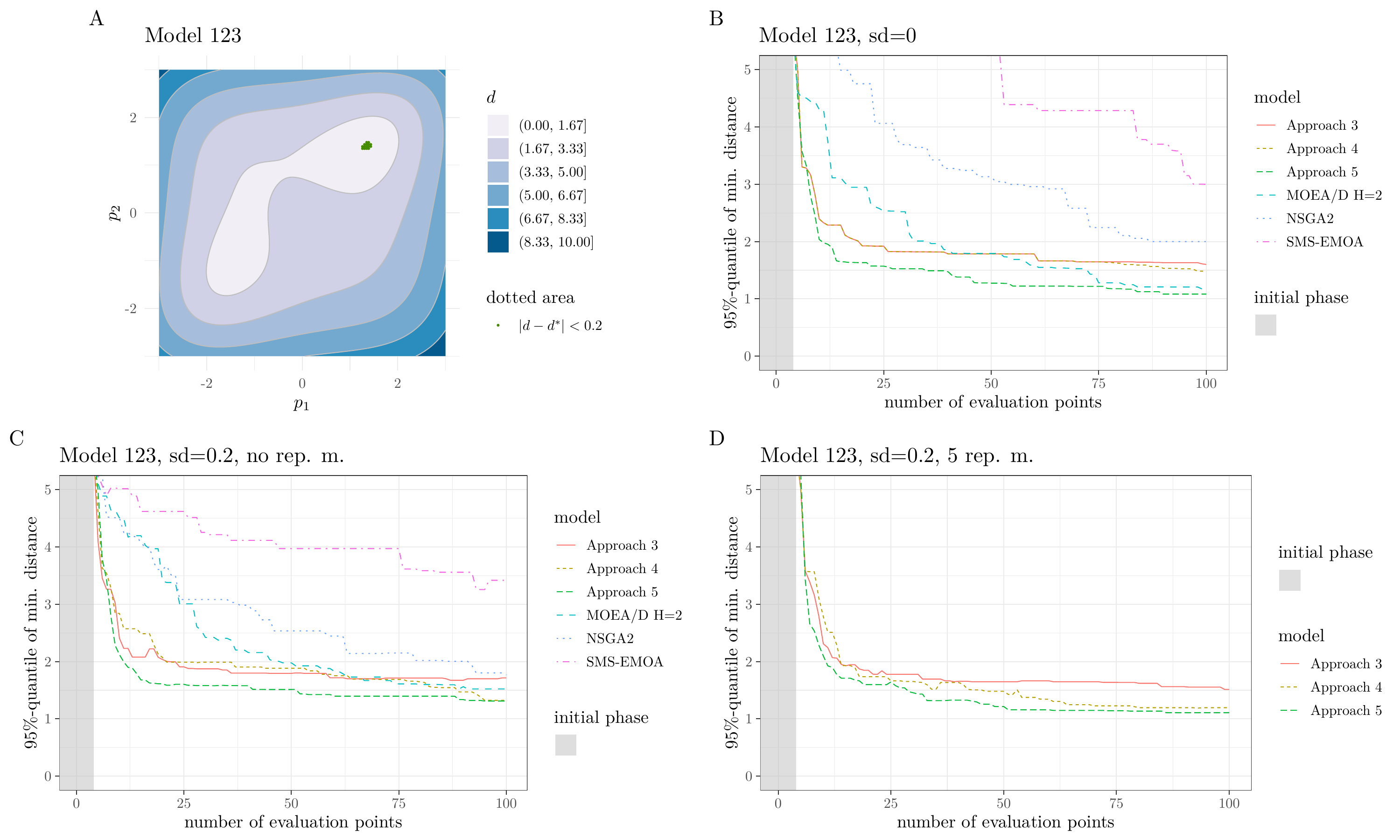} 
\caption{Performances of the approaches 3 -- 5, MOEA/D, NSGA2 and SMS-EMOA for model 123 in case of exact measurements without and with standard deviation.} \label{fig:mod123compare}
\end{figure}

In the same way a bivariate 3-response model \(\phi: \mathbf{p} \mapsto \mathbf{d}, \, \mathbb R^2 \to \mathbb R^3\) is built using components defined by the univariate models 1, 2 and 3, cf.~figure \ref{fig:mod123compare}. The resulting model has one point as exact solution with a surrounding target area. There is a second area with descriptors close to the target, but there is no exact solution in that area.
In the simulations depicted in figures \ref{fig:mod12compare} and \ref{fig:mod123compare} approach 5 turns out to be superior to the considered alternatives with respect to the accuracy of the optimization when \(100\) model evaluations or fewer are considered. Approach 4 is observably weaker and falls, e.g., for the simulations for model 123 without standard deviation and for the model 12 simulations with and without standard deviation, behind MOEA/D. Approach 3 as well as NSGA2 and SMS-EMOA tend to be rather weak competitors.

\hypertarget{results-and-discussion}{%
\section{Results and discussion}\label{results-and-discussion}}

In most of the considered model settings, approach 5 performs clearly better than the chosen alternatives concerning the accuracy and number of required measurements.
The presented algorithm achieved competitive and sometimes superior results by exploiting the relationship between explaining and explained variables on a simplified lower dimensional model. However, it requires a much higher computation time. To compute the simulation paths in the plots for models 12 and 123 comparably high computation times were required, cf.~table \ref{tab:comptime}.

It has to be stressed that in the presented models there exists always at least one theoretical solution of the optimization problem. In cases without exact solution the proposed algorithm will probably fall behind the mentioned evolutionary algorithms. This has to be expected, as in the case that the algorithm cannot find a solution for the currently observed principal component in the prediction step, the search space is extended to sparsely explored areas in a simplistic way without considering the actual measurements of the output data. This might be easily improved by including e.g., the evolutionary algorithms, which have been considered as competitors, as a replacement for the current simplistic fallback for the search in the empty space.
Finding solutions for a multidimensional multi-objective optimization problem in an analytical manner is quite involved.
By working with conditional univariate regression models, the curse of dimensionality has been relieved.
Additionally, dimension reduction techniques were applied to work against the curse of dimensionality and to avoid problems due to multicollinearity in the explaining variables.

\begin{table}[tb!]
\centering
\caption{Computation times in seconds for different algorithms for models 12 and 123}
    \begin{tabular}{l|rrrrrr}
    Model & 12    & \multicolumn{1}{r}{12} & 12    & 123   & \multicolumn{1}{r}{123} & 123 \\
    st. deviation    & 0.2   & \multicolumn{1}{r}{0.2} & 0     & 0.2   & \multicolumn{1}{r}{0.2} & 0 \\
    repetitions  & 1     & \multicolumn{1}{r}{5} & 1     & 1     & \multicolumn{1}{r}{5} & 1 \\\hline
    Approach 1 & 522.126 & \multicolumn{1}{r}{660.121} & 523.23 & 552.015 & \multicolumn{1}{r}{665.478} & 547.737 \\
    Approach 2 & 558.112 & \multicolumn{1}{r}{690.535} & 577.702 & 588.1 & \multicolumn{1}{r}{686.204} & 603.16 \\
    Approach 3 & 722.994 & \multicolumn{1}{r}{1000.339} & 763.01 & 761.715 & \multicolumn{1}{r}{1037.243} & 764.002 \\
    Approach 4 & 726.901 & \multicolumn{1}{r}{719.259} & 748.855 & 752.732 & \multicolumn{1}{r}{871.146} & 765.957 \\
    Approach 5 & 673.862 & \multicolumn{1}{r}{950.781} & 871.341 & 729.618 & \multicolumn{1}{r}{997.623} & 785.884 \\
    NSGA 2 & 9.772 & -    & 10.297 & 12.045 & -    & 12.652 \\
    SMS-EMOA & 25.452 & -    & 25.928 & 28.123 & -    & 28.18 \\
    MOEA/D & 20.63 & -    & 20.701 & 230.042 & -    & 208.367 \\
    \end{tabular}%
\label{tab:comptime}
\end{table}

Possible extensions of the algorithm can make use of further regression techniques, e.g., splines, to model a wider range of relationships. Regression discontinuity designs can help to model jumps.
The question how to design a proper stopping rule for validating the suggested process parameters has been left open for future research.
A further open question is how heteroscedasticity can be addressed appropriately.

\hypertarget{acknowledgement}{%
\section{Acknowledgement}\label{acknowledgement}}

The authors gratefully acknowledge the financial support of this
work as part of the Collaborative Research Center SFB 1232 ``Farbige Zustände'' by
the Deutsche Forschungsgemeinschaft (DFG, German Research Foundation) -- project
number 276397488.

\hypertarget{references}{%
\section*{References}\label{references}}
\addcontentsline{toc}{section}{References}

\hypertarget{refs}{}
\begin{CSLReferences}{1}{0}
\leavevmode\vadjust pre{\hypertarget{ref-Bader2019}{}}%
Bader, Alexander, Anastasiya Toenjes, Nicole Wielki, Andreas Mändle, Ann-Kathrin Onken, Axel von Hehl, Daniel Meyer, Werner Brannath, and Kirsten Tracht. 2019. {``Parameter Optimization in High-Throughput Testing for Structural Materials.''} \emph{Materials} 12 (20): 3439. \url{https://doi.org/10.3390/ma12203439}.

\leavevmode\vadjust pre{\hypertarget{ref-Bartroff2014}{}}%
Bartroff, Jay, and Jinlin Song. 2014. {``Sequential Tests of Multiple Hypotheses Controlling Type {I} and {II} Familywise Error Rates.''} \emph{Journal of Statistical Planning and Inference} 153 (October): 100--114. \url{https://doi.org/10.1016/j.jspi.2014.05.010}.

\leavevmode\vadjust pre{\hypertarget{ref-Beume2007}{}}%
Beume, Nicola, Boris Naujoks, and Michael Emmerich. 2007. {``{SMS}-{EMOA}: Multiobjective Selection Based on Dominated Hypervolume.''} \emph{European Journal of Operational Research} 181 (3): 1653--69. \url{https://doi.org/10.1016/j.ejor.2006.08.008}.

\leavevmode\vadjust pre{\hypertarget{ref-Deb2002}{}}%
Deb, K., A. Pratap, S. Agarwal, and T. Meyarivan. 2002. {``A Fast and Elitist Multiobjective Genetic Algorithm: {NSGA-II}.''} \emph{IEEE Transactions on Evolutionary Computation} 6 (2): 182--97. \url{https://doi.org/10.1109/4235.996017}.

\leavevmode\vadjust pre{\hypertarget{ref-Ellendt2018}{}}%
Ellendt, N., and L. Mädler. 2018. {``High-Throughput Exploration of Evolutionary Structural Materials.''} \emph{{HTM} Journal of Heat Treatment and Materials} 73 (1): 3--12. \url{https://doi.org/10.3139/105.110345}.

\leavevmode\vadjust pre{\hypertarget{ref-Guttman1954}{}}%
Guttman, Louis. 1954. {``Some Necessary Conditions for Common-Factor Analysis.''} \emph{Psychometrika} 19 (2): 149--61. \url{https://doi.org/10.1007/bf02289162}.

\leavevmode\vadjust pre{\hypertarget{ref-Husslage2010}{}}%
Husslage, Bart G. M., Gijs Rennen, Edwin R. van Dam, and Dick den Hertog. 2010. {``Space-Filling {L}atin Hypercube Designs for Computer Experiments.''} \emph{Optimization and Engineering} 12 (4): 611--30. \url{https://doi.org/10.1007/s11081-010-9129-8}.

\leavevmode\vadjust pre{\hypertarget{ref-Amaendle2020}{}}%
Mändle, Andreas. 2020. {``{mvTargetOpt}: Multivariate Multi-Objective Optimization. ({R} Package).''} Zenodo. \url{https://doi.org/10.5281/ZENODO.3885311}.

\leavevmode\vadjust pre{\hypertarget{ref-McKay1979}{}}%
McKay, M. D., R. J. Beckman, and W. J. Conover. 1979. {``A Comparison of Three Methods for Selecting Values of Input Variables in the Analysis of Output from a Computer Code.''} \emph{Technometrics} 21 (2): 239. \url{https://doi.org/10.2307/1268522}.

\leavevmode\vadjust pre{\hypertarget{ref-owenOrth}{}}%
Owen, Art B. 1992. {``Orthogonal Arrays for Computer Experiments, Integration and Visualization.''} \emph{Statistica Sinica} 2 (2): 439--52. \url{http://www.jstor.org/stable/24304869}.

\leavevmode\vadjust pre{\hypertarget{ref-Sobolextquotesingle1967}{}}%
Sobol, I. M. 1967. {``On the Distribution of Points in a Cube and the Approximate Evaluation of Integrals.''} \emph{{USSR} Computational Mathematics and Mathematical Physics} 7 (4): 86--112. \url{https://doi.org/10.1016/0041-5553(67)90144-9}.

\leavevmode\vadjust pre{\hypertarget{ref-Tang1993}{}}%
Tang, Boxin. 1993. {``Orthogonal Array-Based {L}atin Hypercubes.''} \emph{Journal of the American Statistical Association} 88 (424): 1392--97. \url{https://doi.org/10.1080/01621459.1993.10476423}.

\leavevmode\vadjust pre{\hypertarget{ref-vijaya2000}{}}%
Vijayakumar, Sethu, and Stefan Schaal. 2000. {``Locally Weighted Projection Regression: An {O}(n) Algorithm for Incremental Real Time Learning in High Dimensional Space.''} \emph{Proceedings of the Seventeenth International Conference on Machine Learning (ICML 2000)} Vol. 1 (May).

\leavevmode\vadjust pre{\hypertarget{ref-Winkel2020}{}}%
Winkel, Munir A., Jonathan W. Stallrich, Curtis B. Storlie, and Brian J. Reich. 2020. {``Sequential Optimization in Locally Important Dimensions.''} \emph{Technometrics} 63 (2): 236--48. \url{https://doi.org/10.1080/00401706.2020.1714738}.

\leavevmode\vadjust pre{\hypertarget{ref-wold66}{}}%
Wold, Herman. 1966. {``Estimation of Principal Components and Related Models by Iterative Least Squares.''} In \emph{Proceedings of the International Symposium on Multivariate Analysis}, edited by Paruchuri R. Krishnaiah, 391--420. Multivariate Analysis 1. Academic Press, New York. \url{https://ci.nii.ac.jp/naid/20001378860/en/}.

\leavevmode\vadjust pre{\hypertarget{ref-Ye2000}{}}%
Ye, Kenny Q, William Li, and Agus Sudjianto. 2000. {``Algorithmic Construction of Optimal Symmetric {L}atin Hypercube Designs.''} \emph{Journal of Statistical Planning and Inference} 90 (1): 145--59. \url{https://doi.org/10.1016/s0378-3758(00)00105-1}.

\leavevmode\vadjust pre{\hypertarget{ref-Zhang2007}{}}%
Zhang, Qingfu, and Hui Li. 2007. {``{MOEA}/{D}: A Multiobjective Evolutionary Algorithm Based on Decomposition.''} \emph{{IEEE} Transactions on Evolutionary Computation} 11 (6): 712--31. \url{https://doi.org/10.1109/tevc.2007.892759}.

\end{CSLReferences}

\newpage

\hypertarget{appendix}{%
\section*{Appendix}\label{appendix}}
\addcontentsline{toc}{section}{Appendix}

\begin{figure}[ht]

\begin{center}\includegraphics[width=0.99\linewidth]{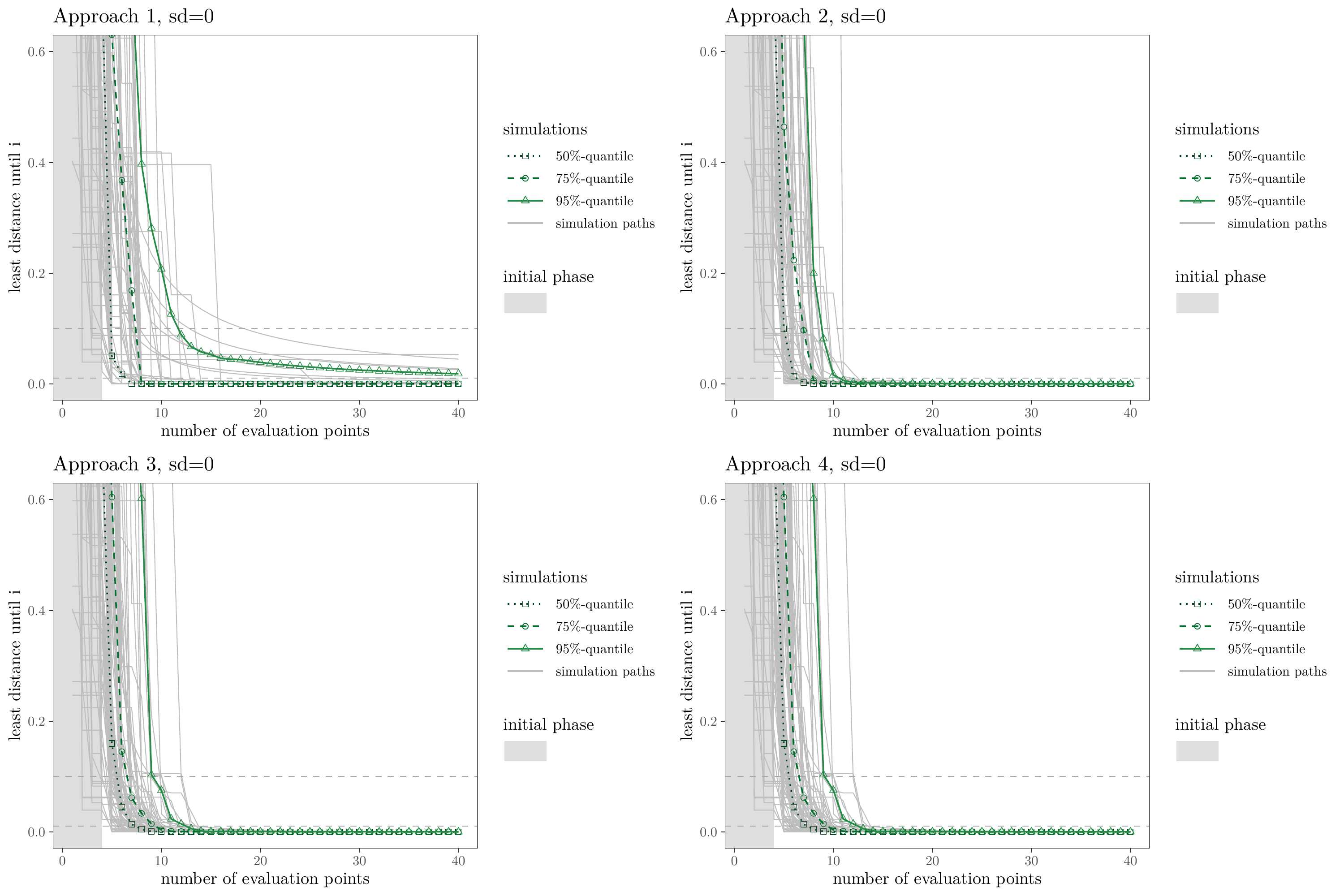} \end{center}
\caption{Performances of the approaches 1 -- 4 for model 1 in case of standard deviation 0 and no repeated measurements.} \label{fig:m1s0plots}
\end{figure}

\newpage

\begin{figure}[h]

\begin{center}\includegraphics[width=0.99\linewidth]{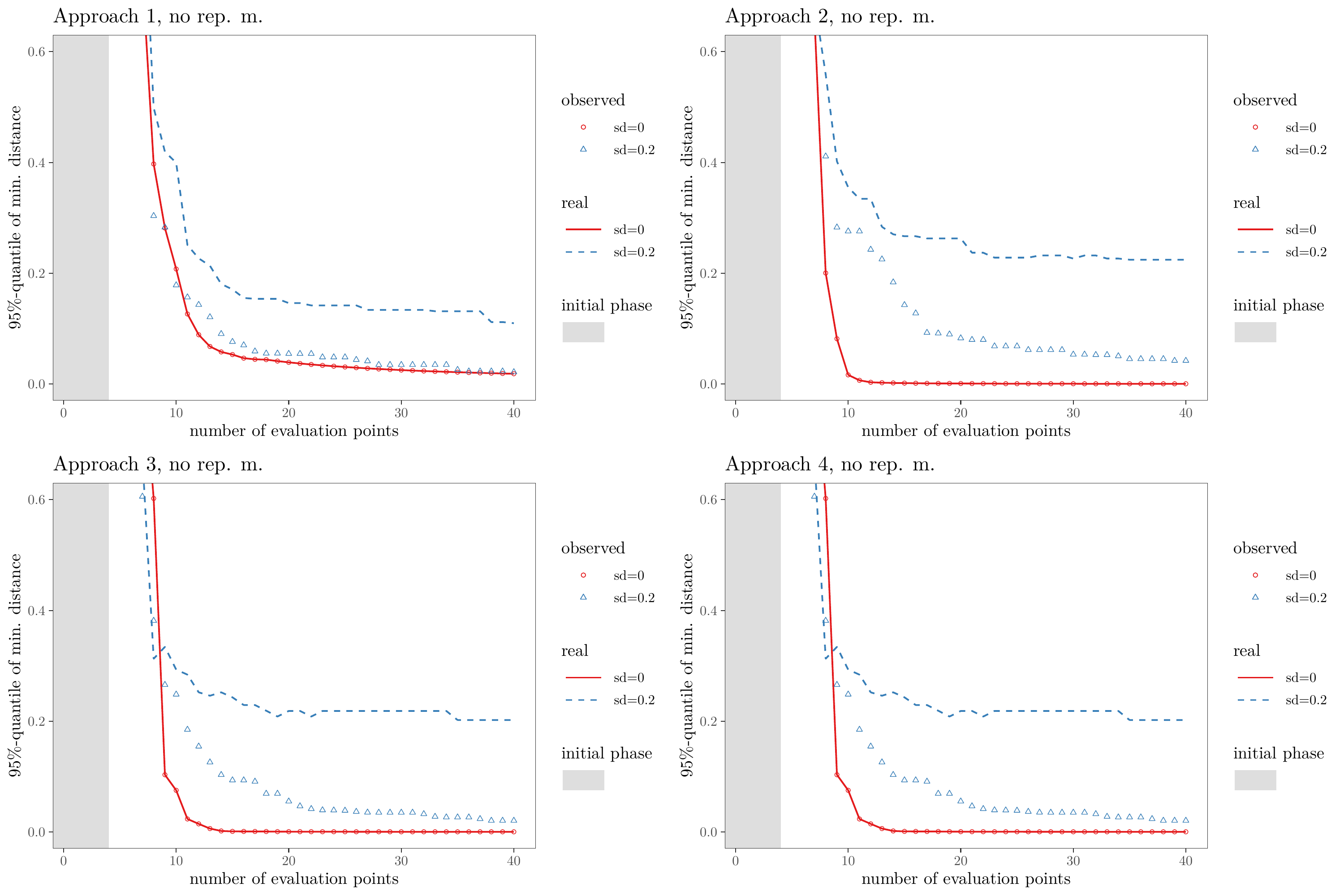} \end{center}
\caption{Performances of the approaches 1 -- 4 for model 1 in case of no repeated measurements.} \label{fig:m1s2rp1plots}
\end{figure}

\newpage

\begin{figure}[h]

\begin{center}\includegraphics[width=0.99\linewidth]{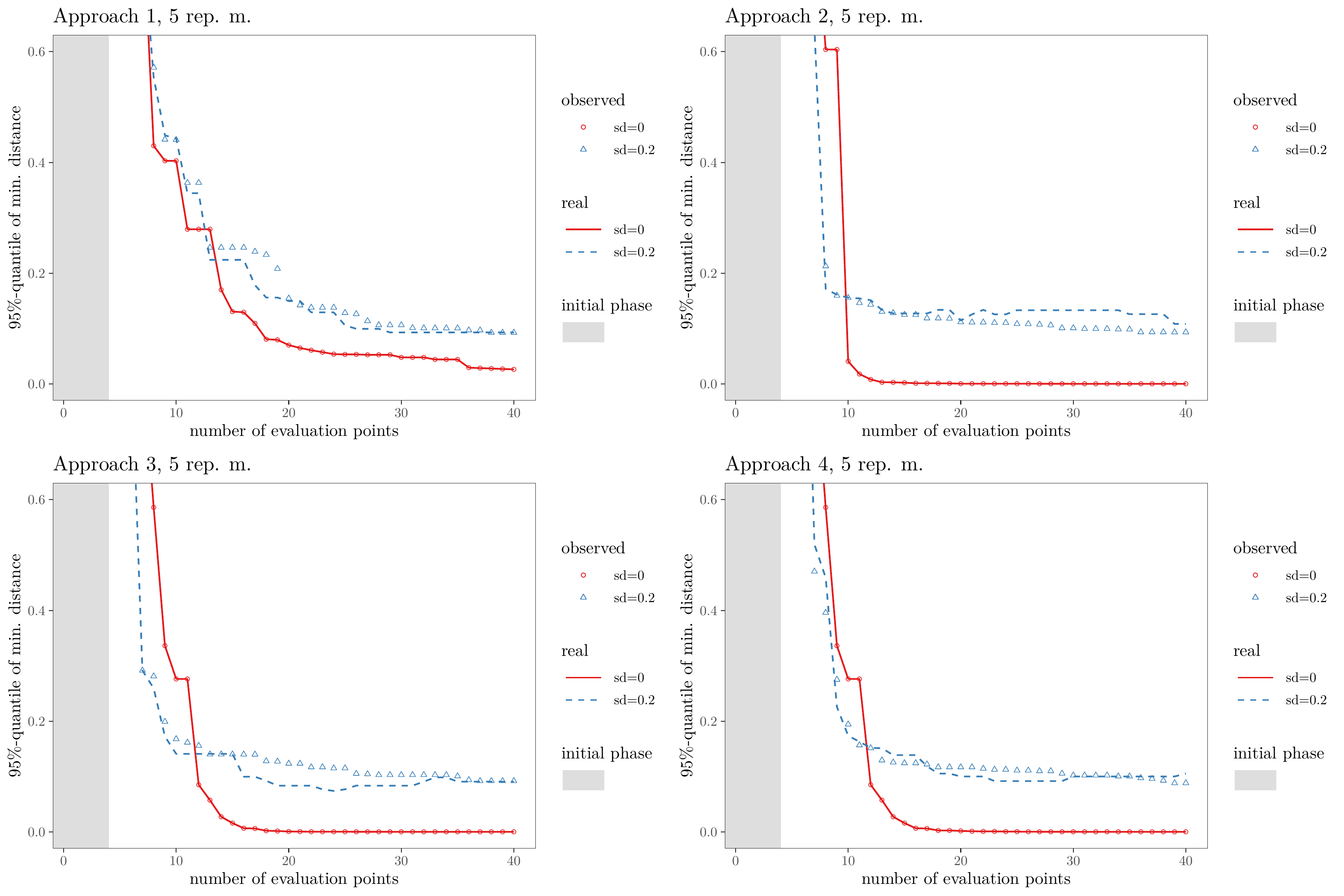} \end{center}
\caption{Performances of the approaches 1 -- 4 for model 1 in case of 5 repeated measurements.} \label{fig:m1s2rp5plots}
\end{figure}

\newpage

\begin{figure}[h]

\begin{center}\includegraphics[width=0.99\linewidth]{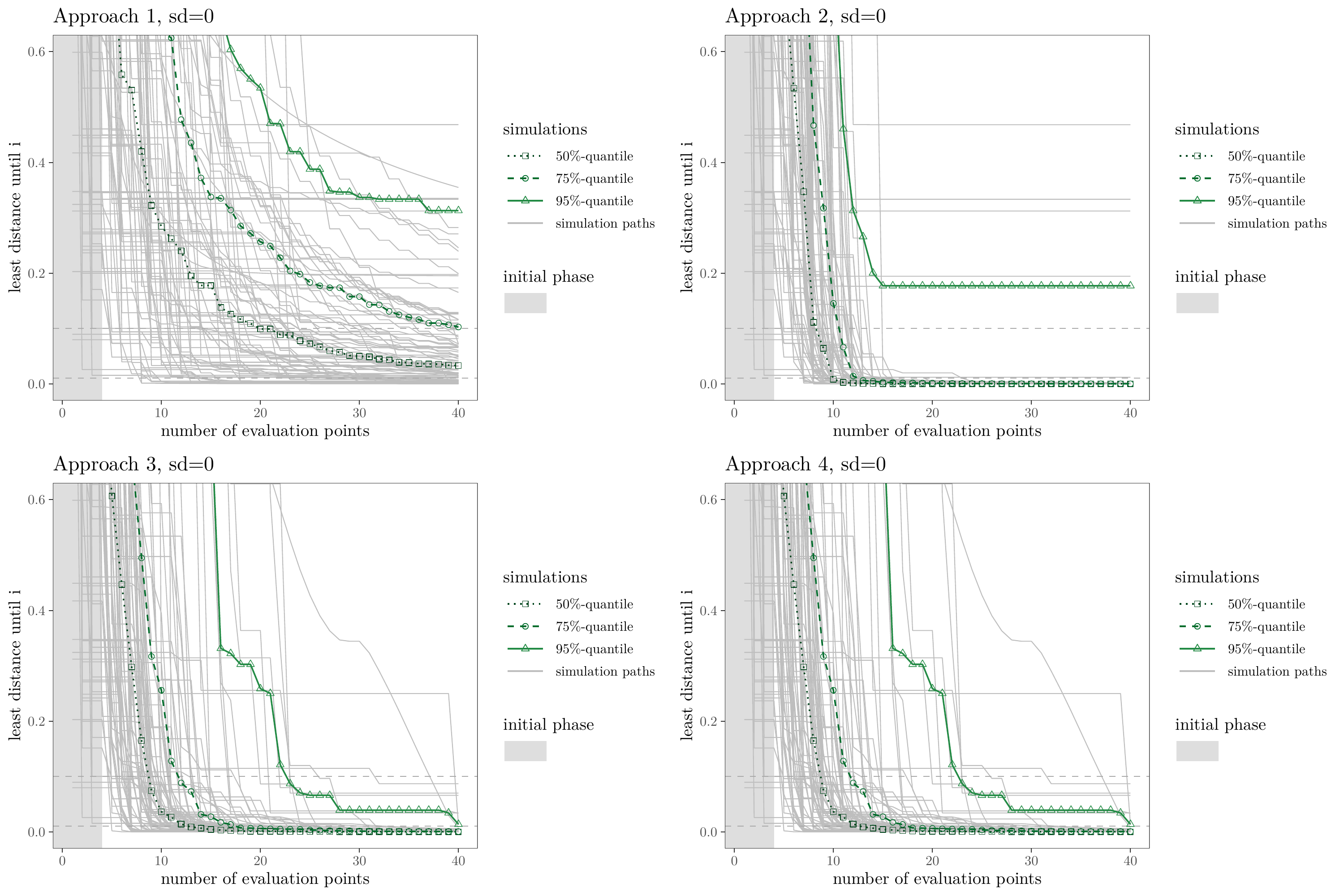} \end{center}
\caption{Performances of the approaches 1 -- 4 for model 2 in case of standard deviation 0 and no repeated measurements.} \label{fig:m2s0plots}
\end{figure}

\newpage

\begin{figure}[h]

\begin{center}\includegraphics[width=0.99\linewidth]{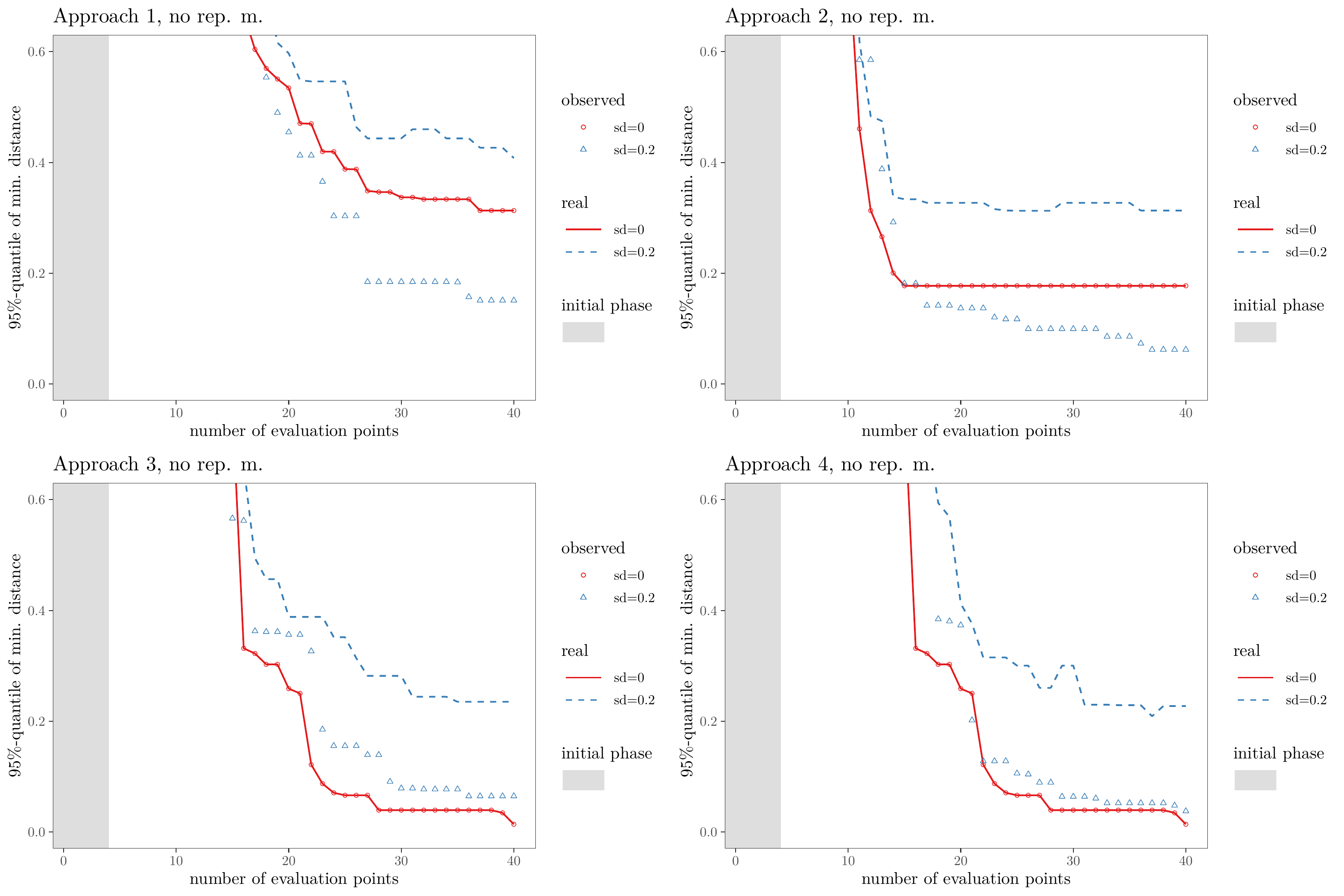} \end{center}
\caption{Performances of the approaches 1 -- 4 for model 2 in case of no repeated measurements.} \label{fig:m2s2rp1plots}
\end{figure}

\newpage

\begin{figure}[h]

\begin{center}\includegraphics[width=0.99\linewidth]{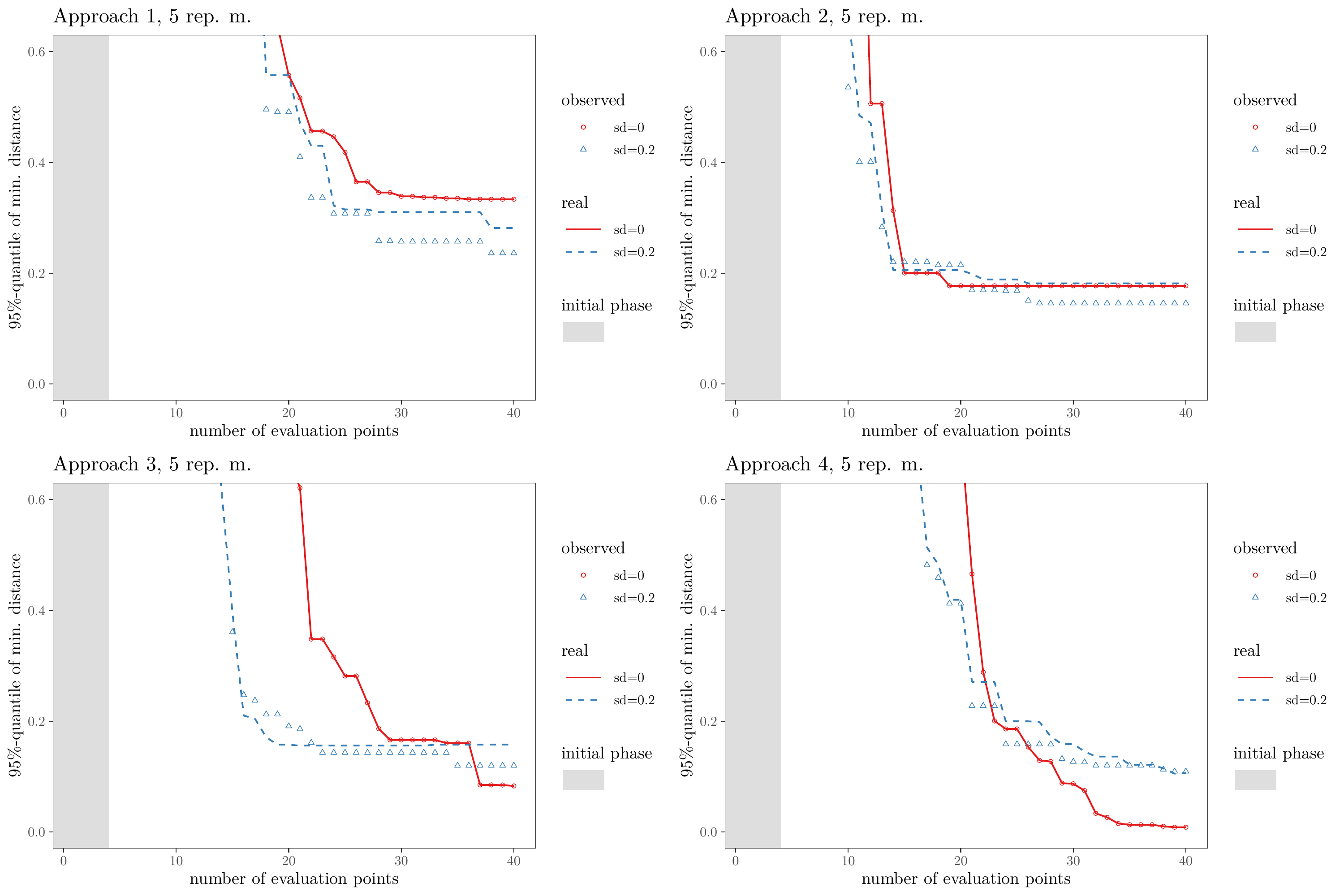} \end{center}
\caption{Performances of the approaches 1 -- 4 for model 2 in case of 5 repeated measurements.} \label{fig:m2s2rp5plots}
\end{figure}

\newpage

\begin{figure}[h]

\begin{center}\includegraphics[width=0.99\linewidth]{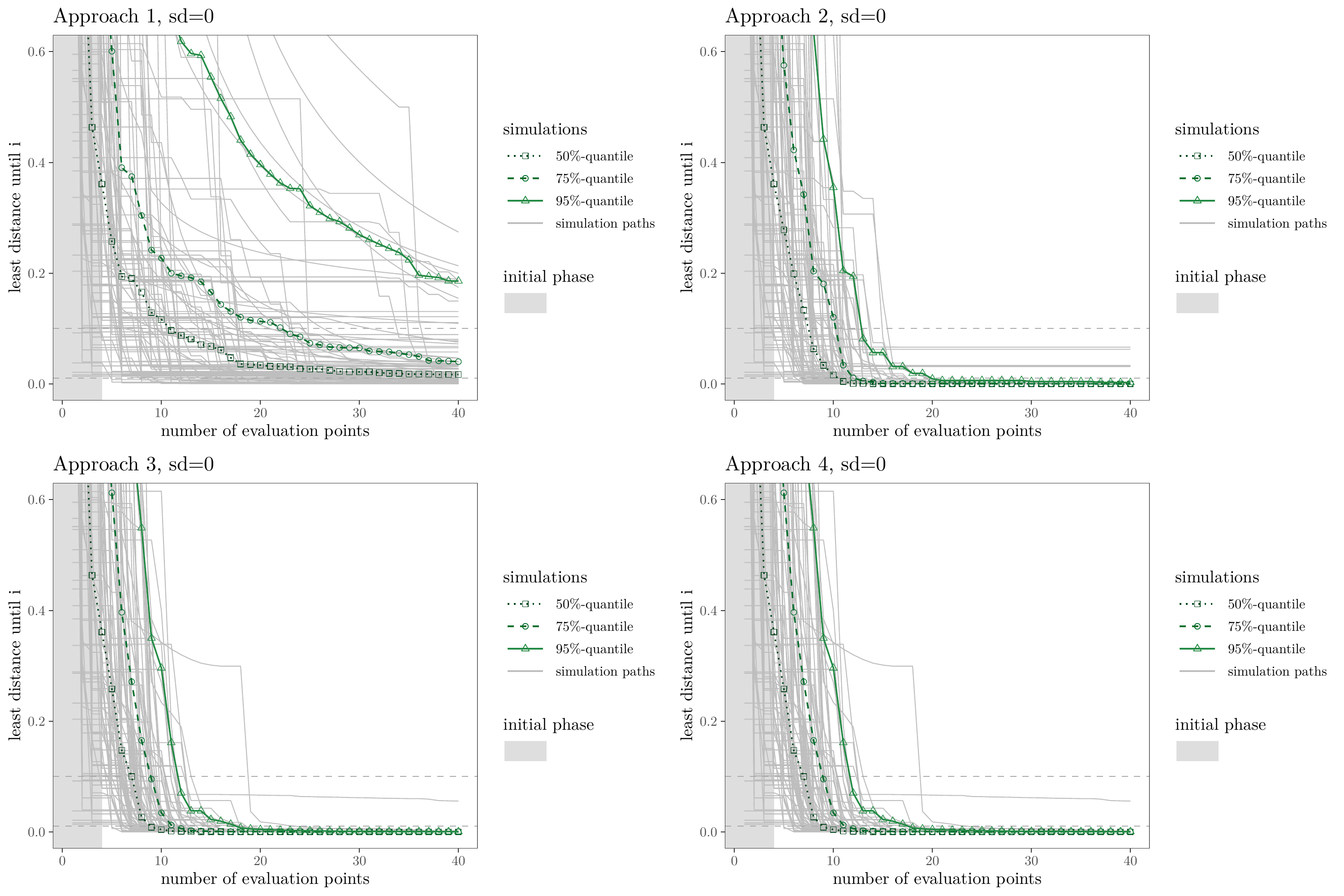} \end{center}
\caption{Performances of the approaches 1 -- 4 for model 3 in case of standard deviation 0 and no repeated measurements.} \label{fig:m3s0plots}
\end{figure}

\newpage

\begin{figure}[h]

\begin{center}\includegraphics[width=0.99\linewidth]{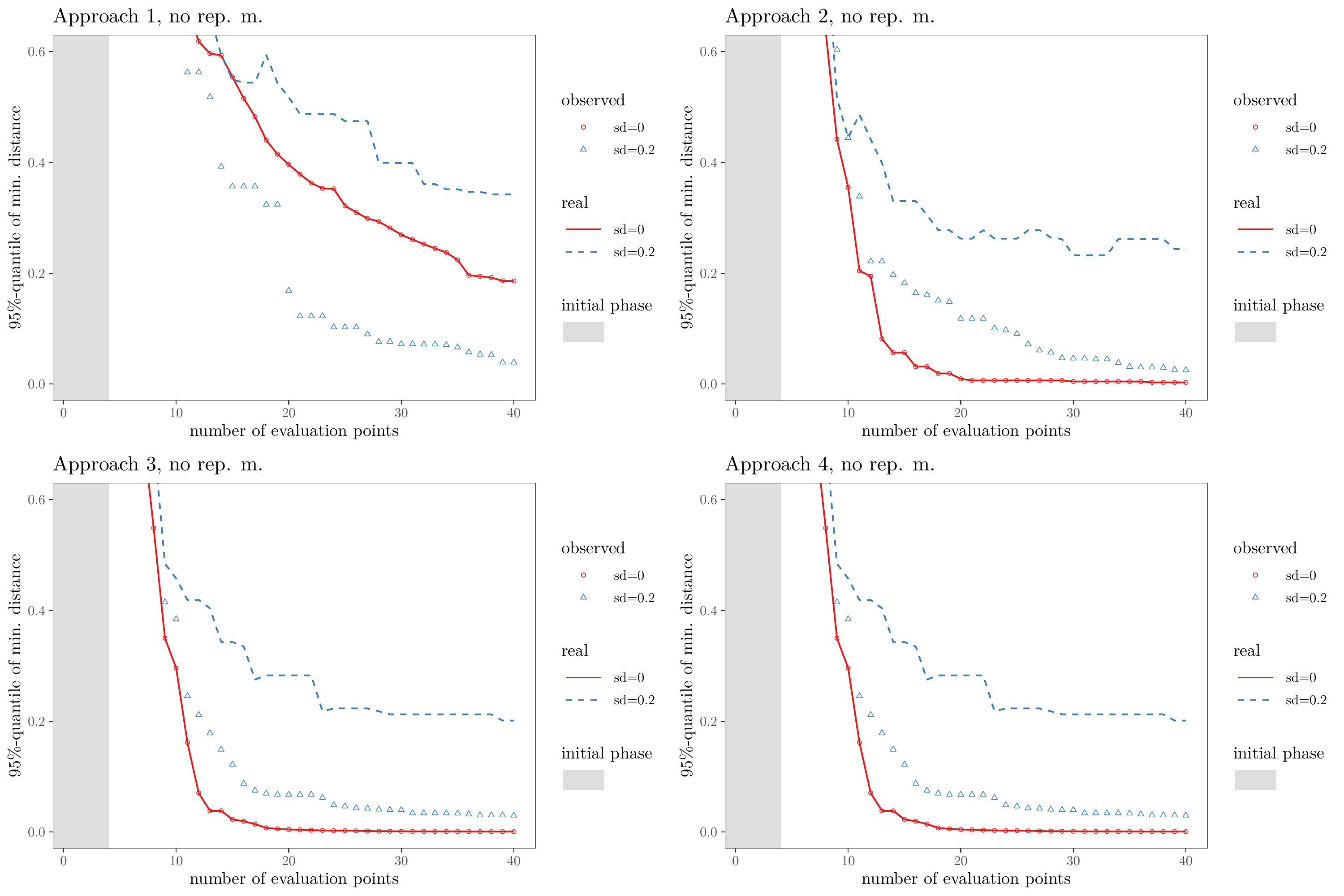} \end{center}
\caption{Performances of the approaches 1 -- 4 for model 3 in case of no repeated measurements.} \label{fig:m3s2rp1plots}
\end{figure}

\newpage

\begin{figure}[h]

\begin{center}\includegraphics[width=0.99\linewidth]{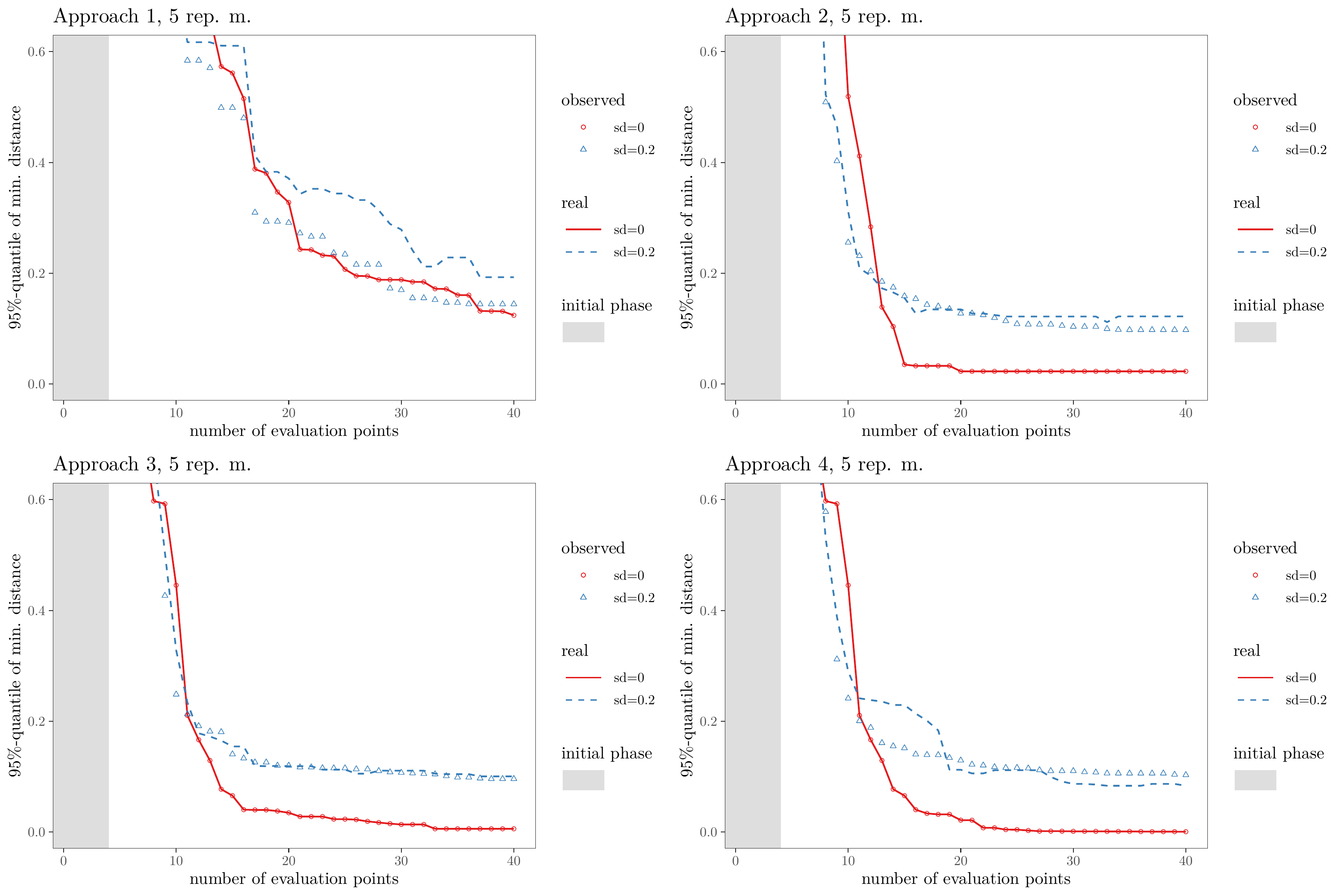} \end{center}
\caption{Performances of the approaches 1 -- 4 for model 3 in case of 5 repeated measurements.} \label{fig:m3s2rp5plots}
\end{figure}

\newpage

\begin{figure}[h]

\begin{center}\includegraphics[width=0.99\linewidth]{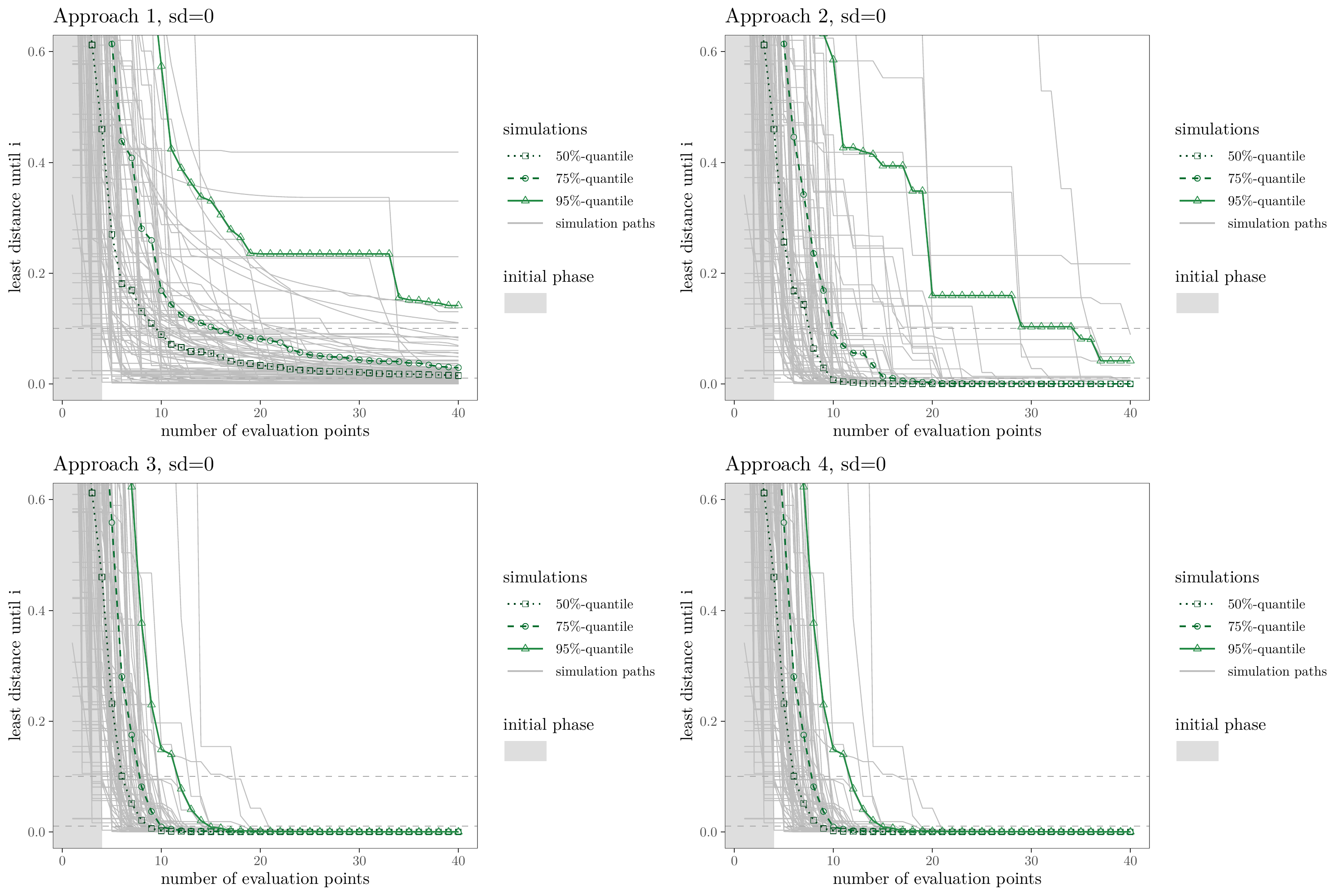} \end{center}
\caption{Performances of the approaches 1 -- 4 for model 4 in case of standard deviation 0 and no repeated measurements.} \label{fig:m4s0plots}
\end{figure}

\newpage

\begin{figure}[h]

\begin{center}\includegraphics[width=0.99\linewidth]{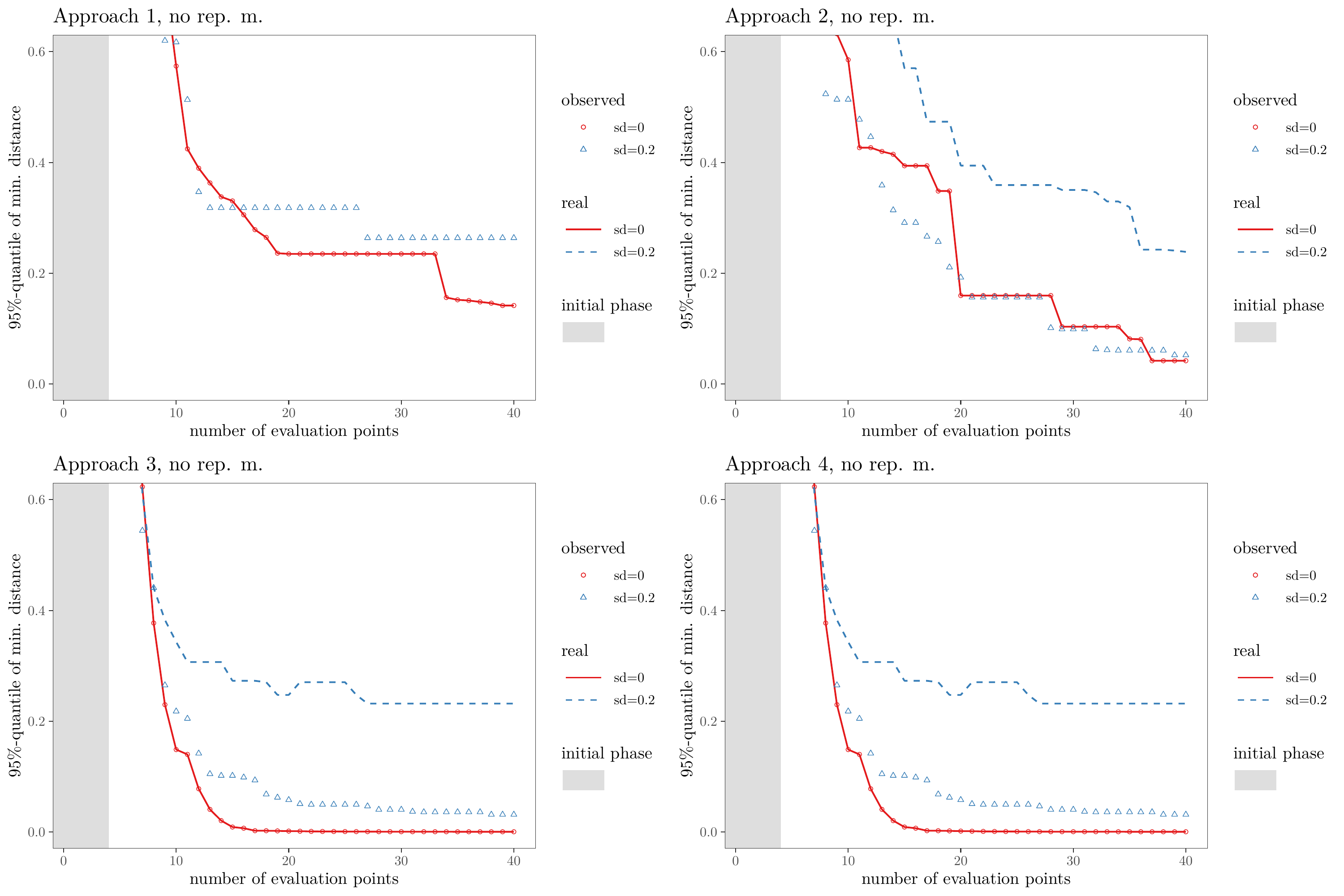} \end{center}
\caption{Performances of the approaches 1 -- 4 for model 4 in case of no repeated measurements.} \label{fig:m4s2rp1plots}
\end{figure}

\newpage

\begin{figure}[h]

\begin{center}\includegraphics[width=0.99\linewidth]{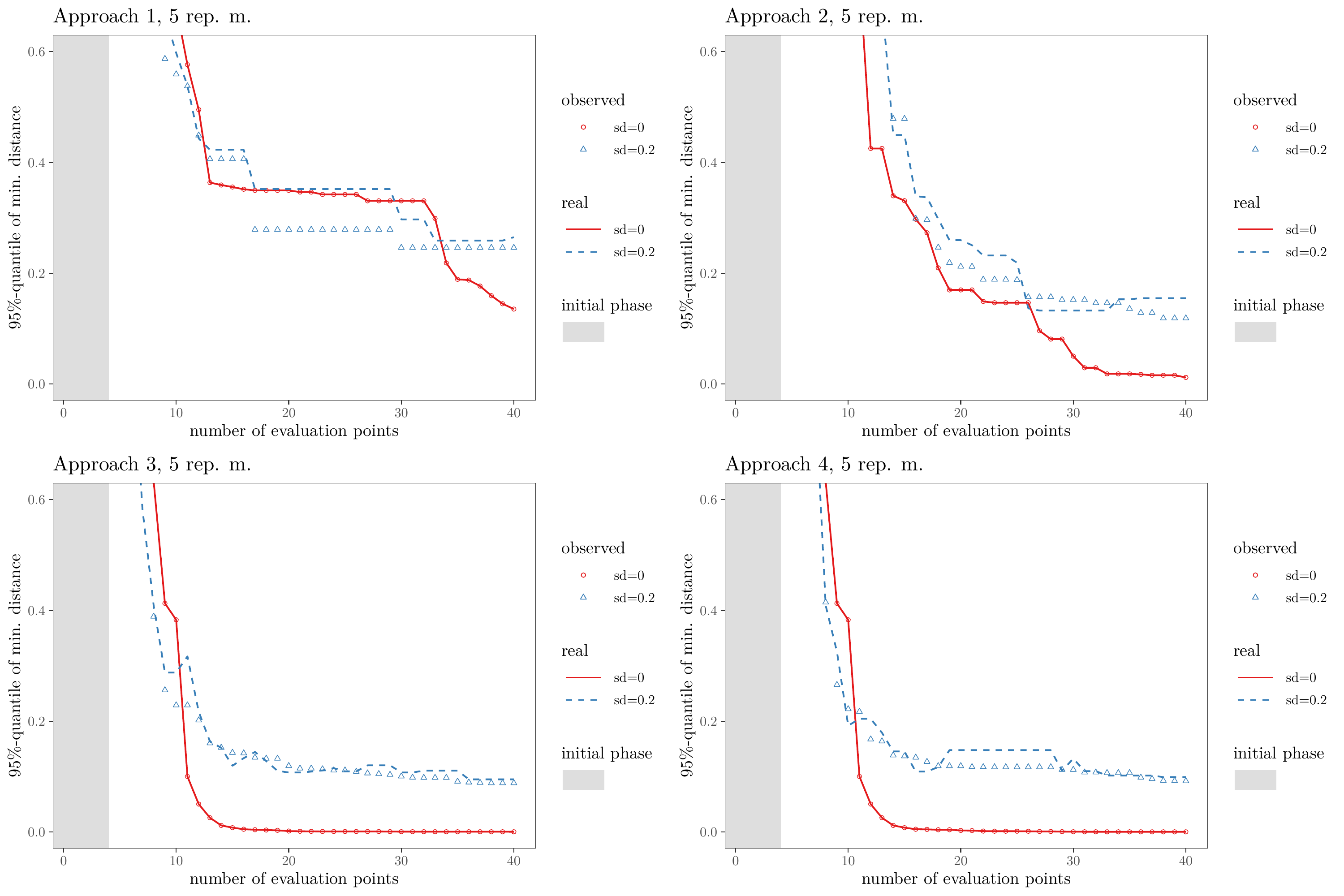} \end{center}
\caption{Performances of the approaches 1 -- 4 for model 4 in case of 5 repeated measurements.} \label{fig:m4s2rp5plots}
\end{figure}

\newpage

\begin{landscape}
\begin{figure}[h]

\begin{center}\includegraphics[width=1\linewidth]{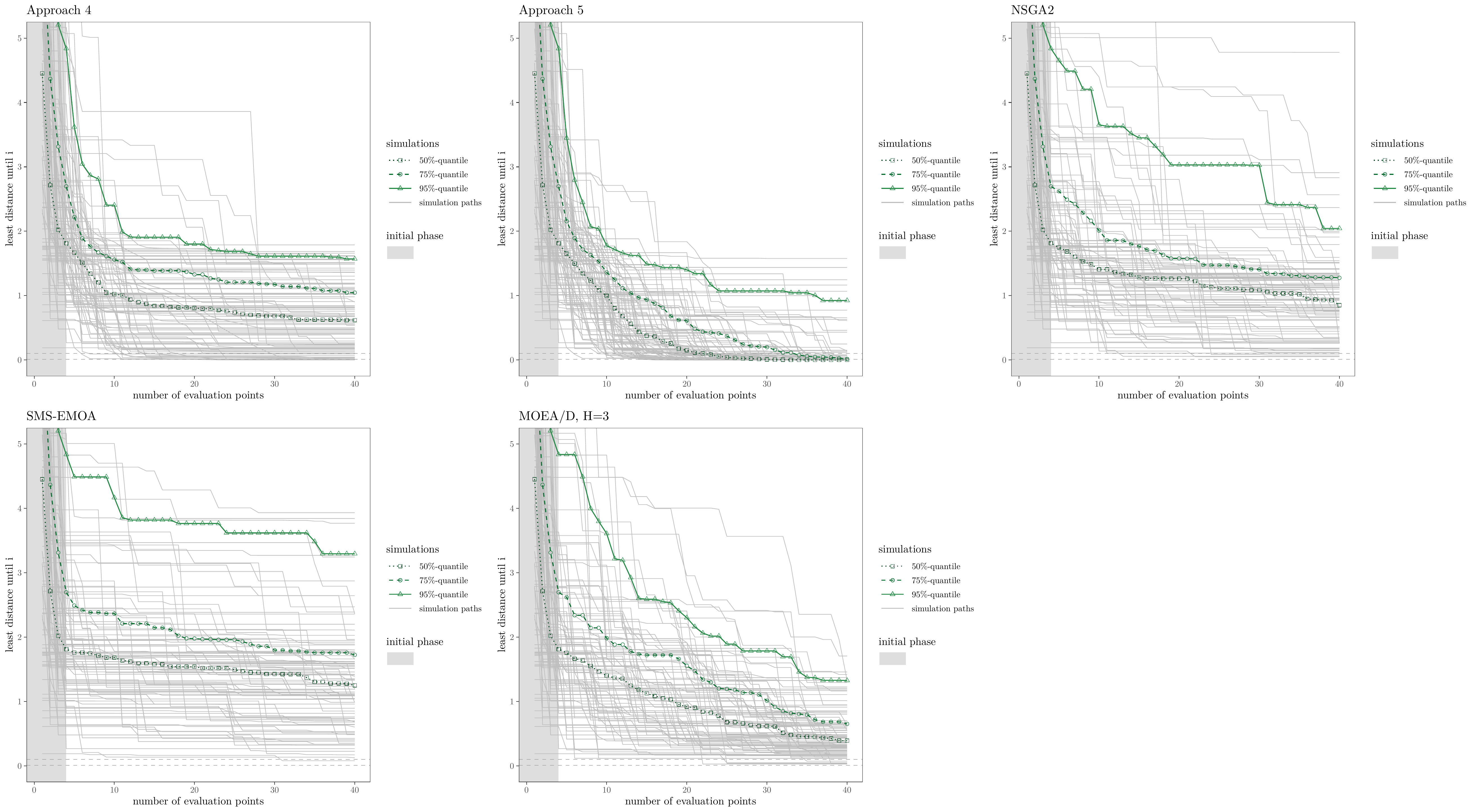} \end{center}
\caption{Performances of the approaches 4, 5, NSGA2, SMS-EMOA and MOEA/D for model 12 in case of standard deviation 0 and no repeated measurements.} \label{fig:mod12moosd0rp0performance}
\end{figure}
\end{landscape}

\newpage

\newpage

\begin{figure}[h]

\begin{center}\includegraphics[width=0.99\linewidth]{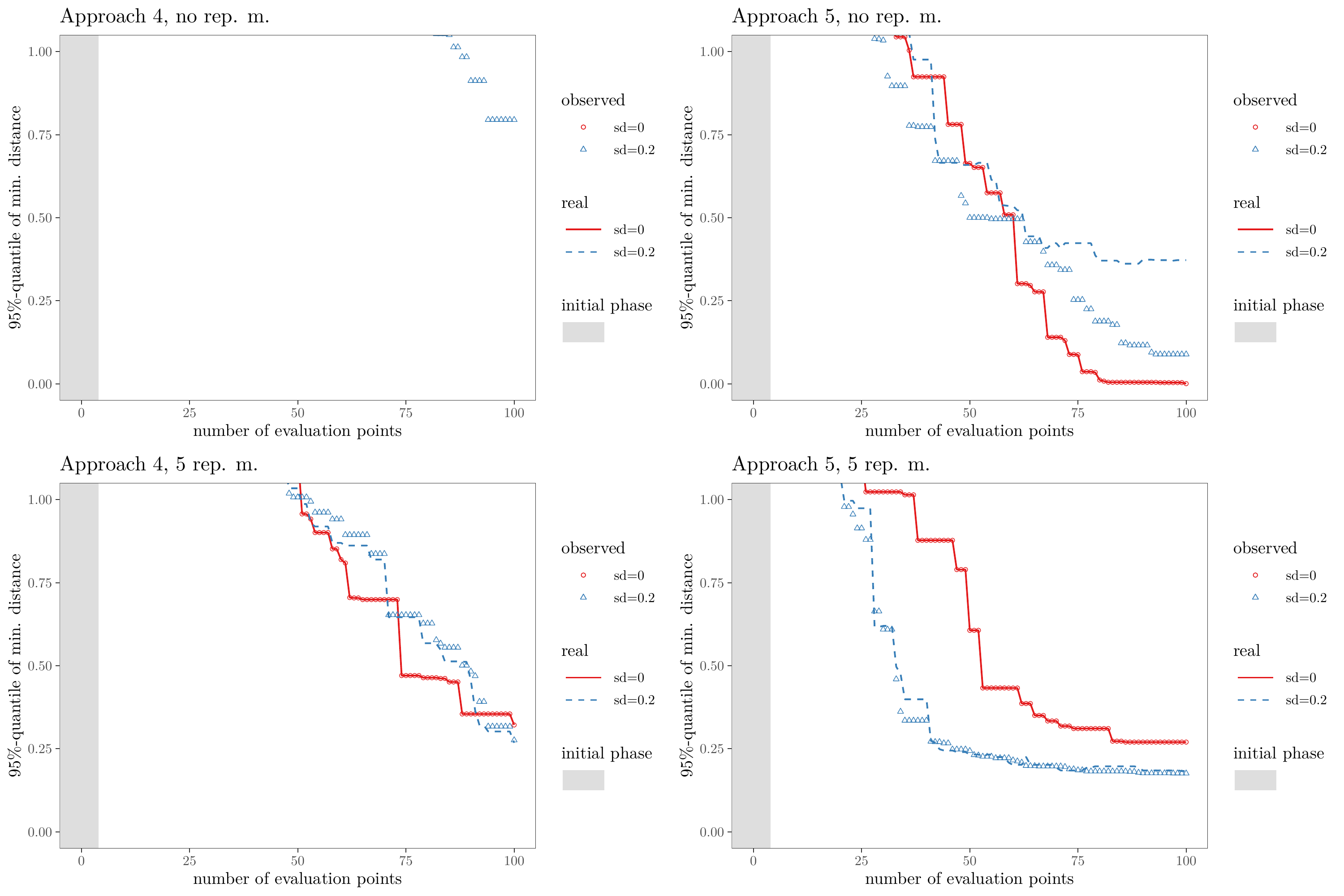} \end{center}
\caption{Performances of the approaches 4 and 5 for model 12 in case of no or 5 repeated measurements.} \label{fig:noname}
\end{figure}

\newpage

\vspace{-1em}

\begin{landscape}
\begin{figure}[h]

\begin{center}\includegraphics[width=1\linewidth]{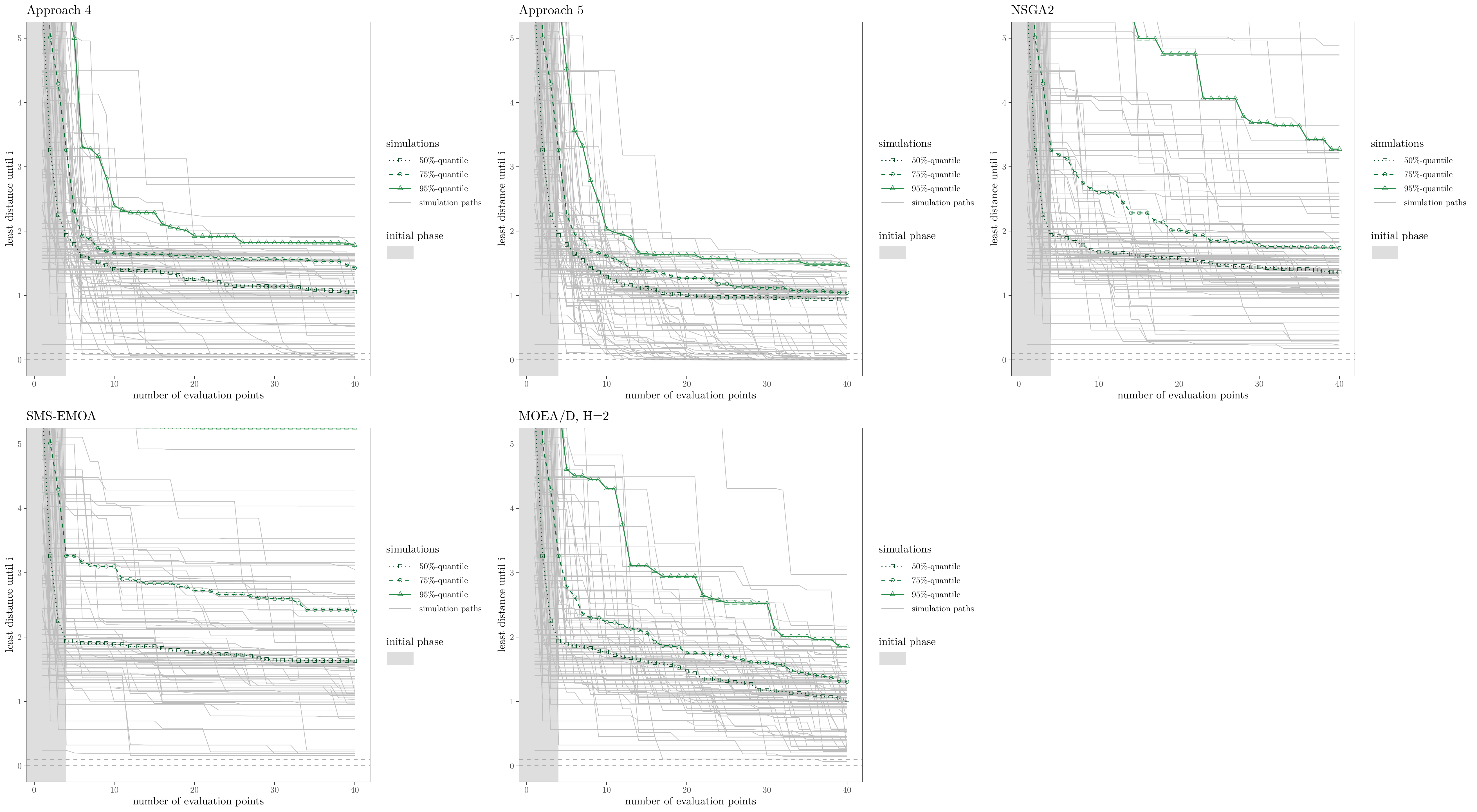} \end{center}

\caption{Performances of the approaches 4, 5, NSGA2, SMS-EMOA and MOEA/D for model 123 in case of standard deviation 0 and no repeated measurements.} \label{fig:mod123ap45moocmparison}
\end{figure}
\end{landscape}

\newpage

\begin{figure}[h]

\begin{center}\includegraphics[width=0.99\linewidth]{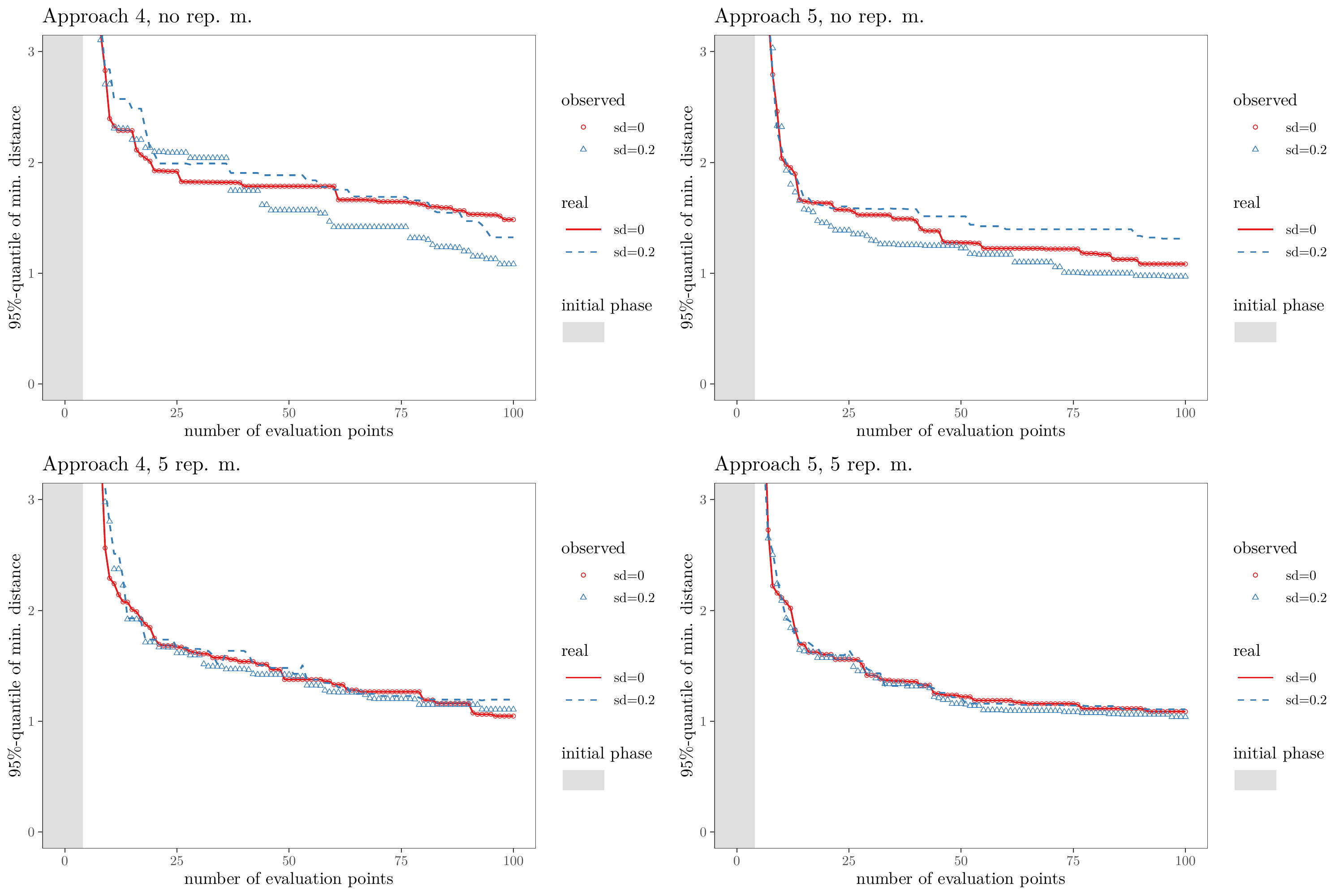} \end{center}
\caption{Performances of the approaches 4 and 5 for model 123 in case of no or 5 repeated measurements.} \label{fig:mod123ap45comparison}
\end{figure}

\bibliographystyle{unsrt}

\end{document}